\documentclass[a4paper,10pt]{article}

\usepackage{a4wide}
\usepackage{graphicx}
\usepackage{subcaption}
\usepackage{url}
\usepackage{color}
\usepackage[usenames,dvipsnames]{xcolor}
\usepackage{amsmath}

\usepackage[normalem]{ulem}

\begin{document}



\title{ GigaGauss solenoidal magnetic field inside of bubbles excited in under-dense plasma }

\author{ Zs. L\'ecz, I. V. Konoplev, A. Seryi, A. Andreev}

\maketitle

Magnetic fields have a crucial role in physics at all scales, from astrophysics to nanoscale phenomena. Large fields, constant or pulsed, allow investigation of material in extreme conditions, opening up plethora of practical applications based on ultra-fast process, and studying phenomena existing only in exotic astro-objects like neutron stars or pulsars. Magnetic fields are indispensable in particle accelerators, for guiding the relativistic particles along a curved trajectory and for making them radiate in synchrotron light sources and in free electron lasers.

In the presented paper we propose a novel and effective method for generating solenoidal quasi-static magnetic field on the GigaGauss level and beyond, in under-dense plasma, using screw-shaped high intensity laser pulses. In comparison with already known techniques which typically rely on interaction with over-dense or solid targets, where radial or toroidal magnetic field localized at the stationary target were generated, our method allows to produce gigantic solenoidal fields, which is co-moving with the driving laser pulse and collinear with accelerated electrons. The solenoidal field is quasi-stationary in the reference frame of the laser pulse and can be used for guiding electron beams and providing synchrotron radiation beam emittance cooling for laser-plasma accelerated electron and positron beams, opening up novel opportunities for designs of the light sources, free electron lasers, and high energy colliders based on laser plasma acceleration.

\section{Introduction}

Magnetic field is one of the fundamental entities which influence nature on all scales. It shapes planets and stars \cite{zweibel97}, electron beams in laser-plasma particle accelerators are guided and focused by magnetic fields \cite{faure04} and it is used to understand natural phenomena under extreme conditions \cite{remington06}. Magnetic field is used to stimulate coherent x-ray radiation from charged particle beams \cite{elias76} and it is responsible for the most spectacular natural phenomena i.e. Aurora Borealis. A capability to generate high strength magnetic field is essential for many projects and research communities (see for example MegaGauss International conferences). The magnetic fields of MegaTesla level strength are observed for example inside neutron stars \cite{spruit98} and can be used to understand some other astrophysical phenomena \cite{nature2012, remington06} as well as matter behavior under extreme conditions \cite{balandina12}. The fields of up to 1~kT strength are currently generated using explosive magnetic generators \cite{chernyshev97} or high-current single shot targets driven by pulsed power generators \cite{lebedev01} which are used to generate high intensity X-ray fluxes. Such machines are capable of generating  magnetic fields of high strength (140~T) \cite{turchi80} and there are prediction of possible generation of up to  600~T in some cases \cite{slough12}. The single shot techniques are well developed and allows generating of fields around 1~kT. Alternatively if non destructive machines are used the maximum strength field achieved using solenoids (without solenoid destruction) is reported to be 100~T at Los Alamos National Laboratory \cite{sims08}. Reaching magnetic fields of a GigaGauss strength levels seems to be impossible as electron currents which is required to drive such fields and $\mathbf{J}\times \mathbf{B}$ self induced forces will destroy any currently available material. 

Developments of laser technology stimulated exploration of methods of generation of large magnetic fields from laser pulses directly or via interaction with plasma or solid targets. Perhaps the most close to conventional methods where pulsed solenoids are used is the method described in \cite{fujioka2013} where kilotesla magnetic field was generated by pulsed laser in a capacitor-coil target configuration. Predictions of large toroidal magnetic field generated by laser pulse in near-critical plasma were made in \cite{puhkov1996} and experimental observations reported in \cite{borghesi1998}, \cite{tatarakis2002} or \cite{sarri2012} demonstrated production of megagauss magnetic fields in laser interaction with solid targets. Note that in all these cases the field is generated in a stationary space volume, the target is either solid or near-critical plasma, and direction of the field is radial or toroidal, i.e. not solenoidal and not quasi-static in co-moving electron beam frame. 

More recently, generation of longitudinal solenoidal magnetic field was explored, based on use of laser pulses carrying orbital angular momentum (OAM) created either by polarization of the laser pulse \cite{ali2010,plasmaRot,amplify} or by a hollow screw-like intense Laguerre–Gaussian laser pulse \cite{wang2015}. These techniques are closer to that we discuss in our paper, however method we suggest is capable to produce significantly higher solenoidal fields. 

The way of generation of high amplitude solenoidal field generation, suggested and considered in detail in this paper, is based on interaction of screw-shaped laser pulse\cite{plasmaRot} interacting with under-dense plasma. This interaction creates magnetic fields of GigaGauss strength level inside a volume of $0.1 - 10s$ $\mu$m transverse dimensions, depending on the plasma density and pulse spirality, co-moving with the laser pulse, and thus suitable for effective interaction with the accelerated electron or positron beams. We no that the generation of the solenoidal field can also be achieved by using charged particle (electron) beams, but this will be considered in the following studies. 

Generation of a screw-shaped laser pulse is an interesting topic in its own right \cite{amplify}. We assume that it can be generated via compression/focusing along a single rotating plane of circular polarized beam or shaping it with relativistic electrons \cite{attospiral, helical} and amplifying the generated higher harmonics \cite{amplify}, and will focus on utilization of such a screw-shaped laser pulse. 

In this paper two operating regimes will be discussed under which magnetic field can be generated by the screw-shaped laser pulse, namely: bubble solenoid (short solenoid) and steady solenoid (long solenoid) regimes. In the first regime the solenoidal magnetic field is generated by the electron currents moving around plasma bubble shell in a similar to solenoid manner with the highest field observed at the end of the bubble where electrons’ trajectories are collapsing creating high strength longitudinal magnetic field. In this case the field's strength is adiabatically changing from the center of the bubble to its end reaching the highest amplitude. The magnetic field direction coincides with the direction of laser propagation with the volume (few cubic micrometers) having the highest field, defined by the step of the laser spiral. The second regime is realized when a set of conditions (discussed below in the paper) are such that the plasma bubble is not formed and the solenoid currents are not limited to the dimensions of the plasma wavelength. In this case the field can be sustained over longer distances, until the pulse energy is depleted, but the field amplitude achieved in this regime will be order of magnitude less (i.e. $10s$ kT).

The paper is structured as follow: the second section is dedicated to the model description while in the third section different regimes will be considered. We discuss the application of the bubble solenoid regime to laser plasma acceleration of electron or positron beams at the end of the second section, demonstrating the fast synchrtotron radiation cooling of the emittance of the accelerating beams. In the conclusion we will discuss the results and outline future work and possible impacts. One notes that the capability to generate such fields will allow dramatic progress in non-destructive, reproducible studies of phenomena at fields strength which are not yet accessible in the laboratories. In this work we focus in particular on the novel opportunities opened for design of laser-acceleration based devices -- light sources, free electron lasers and high energy colliders.

\section{Electron blow-out and collapse}

The laser-plasma (LP) interaction is one of the most dynamic research areas which is driven by number of problems including design of compact light sources and particle accelerators. Another challenge, which can be resolved using LP interaction is generation of magnetic fields of MegaTesla (MT) strength. One notes that to resolve this problem the laser-plasma parameters should be different from ones used in conventional studies \cite{review} and in general may not be optimal for example for LP acceleration of electrons. 

Let us consider the condition $\lambda_p\gg \lambda_L$, where $\lambda_p=2\pi c/\omega_p$ is the plasma wavelength with $\omega_p=(n_0 e^2/(\gamma m_e \epsilon_0))^{1/2}$ (relativistic plasma frequency, where $n_0$ is the unperturbed plasma density, $\gamma$ and $m_e$ is the relativistic factor and mass of electron) and $\lambda_L$ is the laser wavelength. In this case the average ponderomotive force of the pulse, which is proportional to the gradient of the intensity defined by the envelope function, acts on the electrons \cite{envelopeModel, Gordon2000}. This is well known envelope model which is widely used for investigation of laser wake field acceleration (LWFA) in low density plasmas \cite{PMessmer}. This is appropriate and valid "short-cut" to minimise the spatial resolution of numerical modeling which leads to reduction of calculation time. We use a similar model and the same concept for studying the effect of laser pulse shape and its angular momentum on the bubble formation and magnetic field generation. We note that the 3D models of these phenomena have been studied using 3D Particle in Cell (PIC) code VSim, which has a self-consistent light-frame envelope model implemented based on the method from Ref. \cite{PMessmer}.

\begin{figure}[h]
\centering
\begin{subfigure}[b]{0.3\textwidth}
  \includegraphics[trim = 0mm -50mm 0mm 20mm,width=\textwidth]{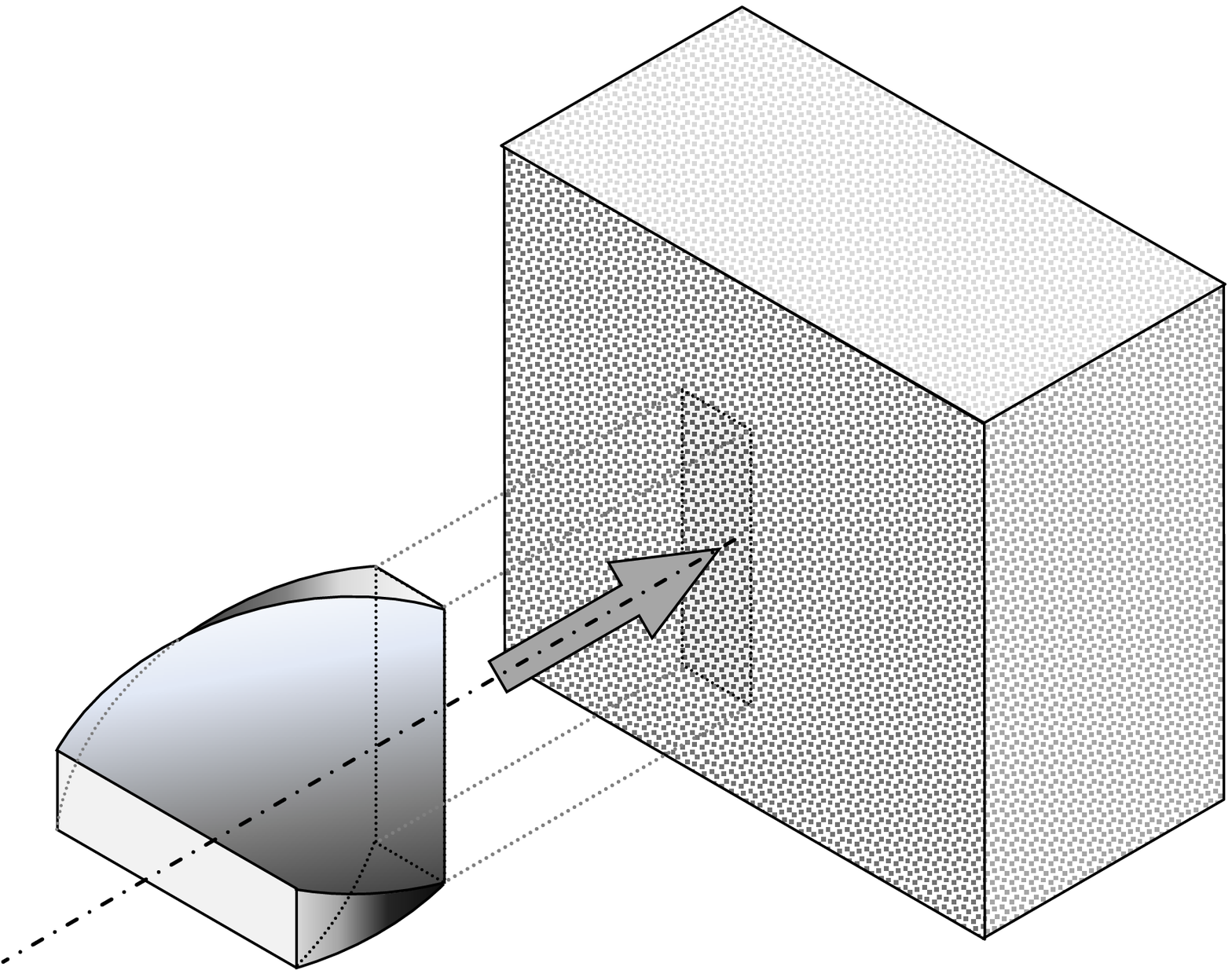}
  \caption{}
\end{subfigure}
\begin{subfigure}[b]{0.6\textwidth}
  \includegraphics[width=\textwidth]{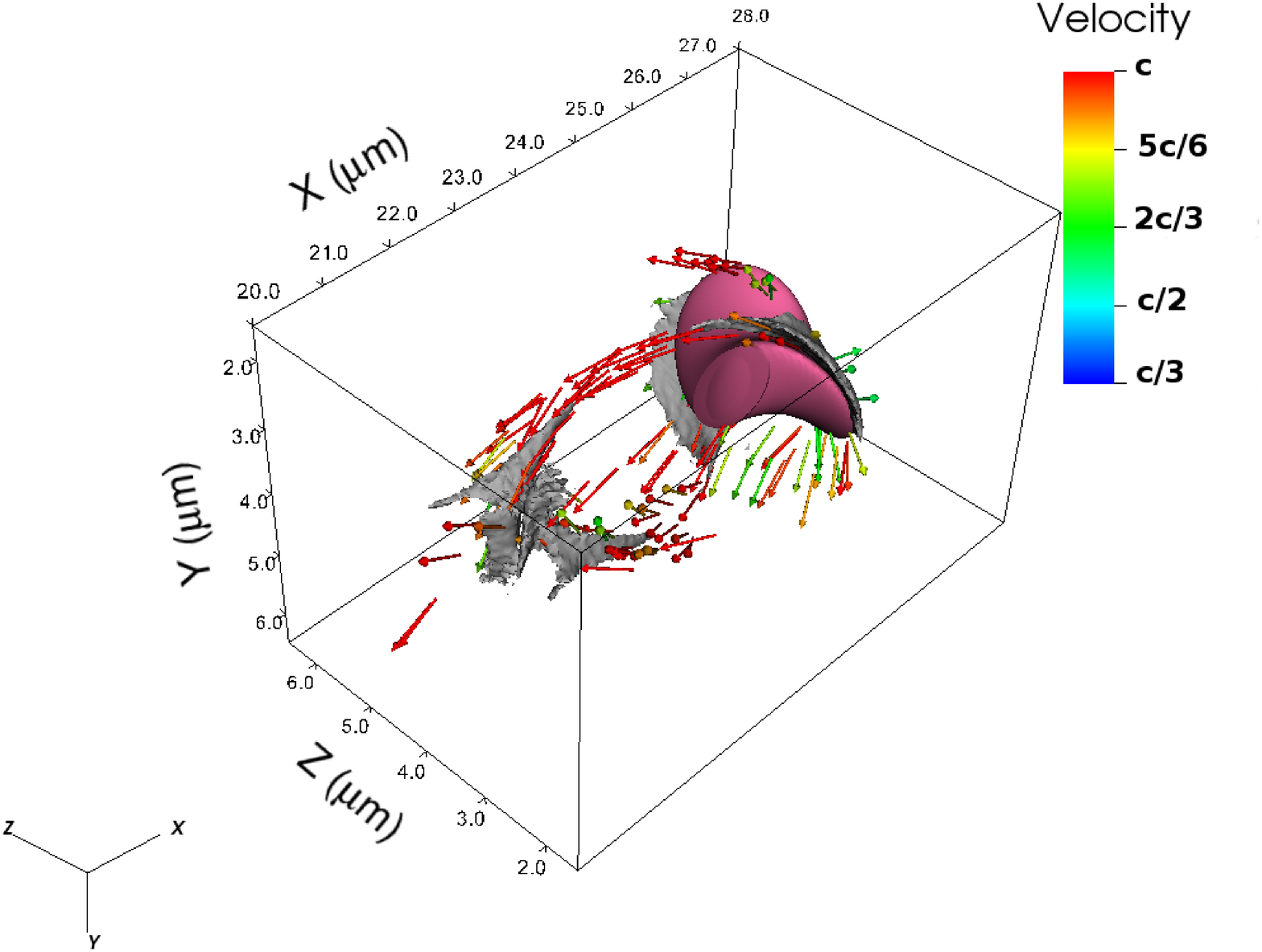}
  \caption{}
\end{subfigure}  
\caption{ {\bf (a)} Schematic of the laser pulse moving into the plasma (grey box) with momentary projection of the pulse front on the box. {\bf (b)} illustrates the isosurface of pulse intensity (purple) at the value of $2\times 3\times 10^{20}$ W/cm$^2$ and electron density (gray) with $10^{26}$ m$^{-3}$. The arrows represent the velocity vectors of electrons seen from the moving frame of the laser pulse and the color code shows their magnitude.  }
\label{isosurf_elvec}
\end{figure}

The exact analytical description of the laser pulse is presented in the Appendix. The laser pulse shape i.e. its envelope function, illustrated in Fig. \ref{isosurf_elvec}a, is similar to a drill bit propagating along x-coordinate and Fig. \ref{isosurf_elvec}b shows the electron dynamics while the pulse moves through the plasma. The intensity distribution of the pulse is Gaussian-like, but it is modulated helically, providing the desired spiral shape. One notes that the main difference (in plasma dynamics) between the standard Gaussian and the spiral-shaped beam propagation through the plasma is the azimuthal non-uniformity of electron density (i.e. appearance of azimuthal current) along the bubble surface.

The electrons expelled by the laser pulse will move along spiral trajectories on the surface of the generated bubble (azimuthal electron current), which in turn induces strong axial magnetic field with the maximum value in the back of the bubble. Indeed the spiral shaped laser pulse gives a twist to the electrons which can be seen at the right side of Fig. \ref{isosurf_elvec}b (in the vicinity of the pulse). As they travel to the back side of the bubble two electric current channel are formed, which approach each other and finally pass by within a distance less then laser spot diameter. The currents cannot merge as these two electron currents have opposite direction, repelling each other and thus limiting the magnetic field generated at the merging point. However, the maximum value of the field is generated at the end of the bubble due to the fact that the electrons' trajectories are approaching each other in this region. This motion as well as dense electron bunches are shown in Fig. \ref{isosurf_elvec}b. To minimize the calculation time and for better illustration of the electron currents we move to the  frame co-moving with the laser pulse. We also note that throughout the paper the laser wavelength is adjusted such that the conditions $\lambda_L\ll \lambda_{sp}$, where $\lambda_{sp}$ is the spiral step, and $n_0<n_{cr}$, where $n_{cr}=\omega_L^2 m_e\epsilon_0/e^2$, are fulfilled. 

In the example shown in Fig. \ref{isosurf_elvec}b the peak intensity is $I_0=1.6\times 10^{21}$ W/cm$^2$, the pulse width is $1.8 \mu$m (FWHM), the pulse length is equal to the spiral step, $\lambda_{sp}=1.8 \mu$m, and electron density is typical fully ionized C-H gas density, $n_0=0.62\times 10^{-3}n_{cr}=7\times 10^{25}$ m$^{-3}$. In order to use the envelope model the plasma has to be strongly underdense, i.e. $n_0/n_{cr}\ll 1$, to ignore the pulse energy depletion during the short time scale considered here. The pulse evolution during propagation in underdense plasma is not trivial \cite{pulseEvol}. The exact pulse and field evolution over a longer time is out of scope of this work, but these aspects will be studied in the future. Although the plasma is dilute the pulse shape can be strongly modified near the wings of the radial Gaussian profile, where the edge of the helical intensity modulation is trimmed by the repelled electrons. This reduces the spirality of the pulse and results in a much smoother quasi-Gaussian beam. On the other hand the spiral step can decrease due to pulse steepening \cite{review} at the front side of the laser pulse, which will change the field distribution in the back of the bubble. All these effects make the interaction extremely complicated, therefore we limit ourselves to a narrow parameter-range and time window, where the generated B-field is quasi-static. Later in this paper we show examples, where the pulse evolution can not be neglected even on this short time scale.

\begin{figure}[h]
\centering
\begin{subfigure}[b]{0.55\textwidth}
\includegraphics[width=\textwidth]{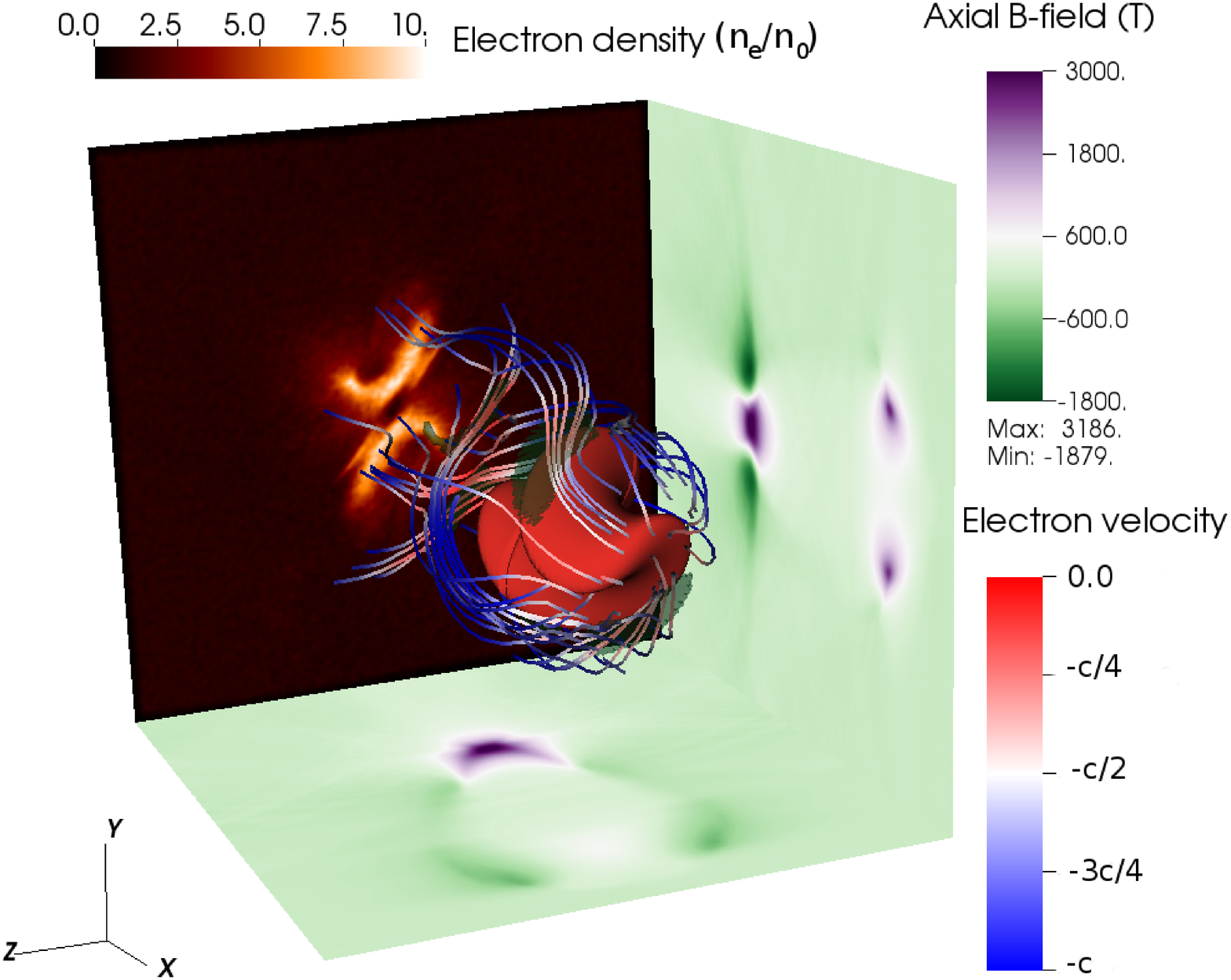}
  \caption{}
\end{subfigure}
\begin{subfigure}[b]{0.40\textwidth}
\includegraphics[width=\textwidth]{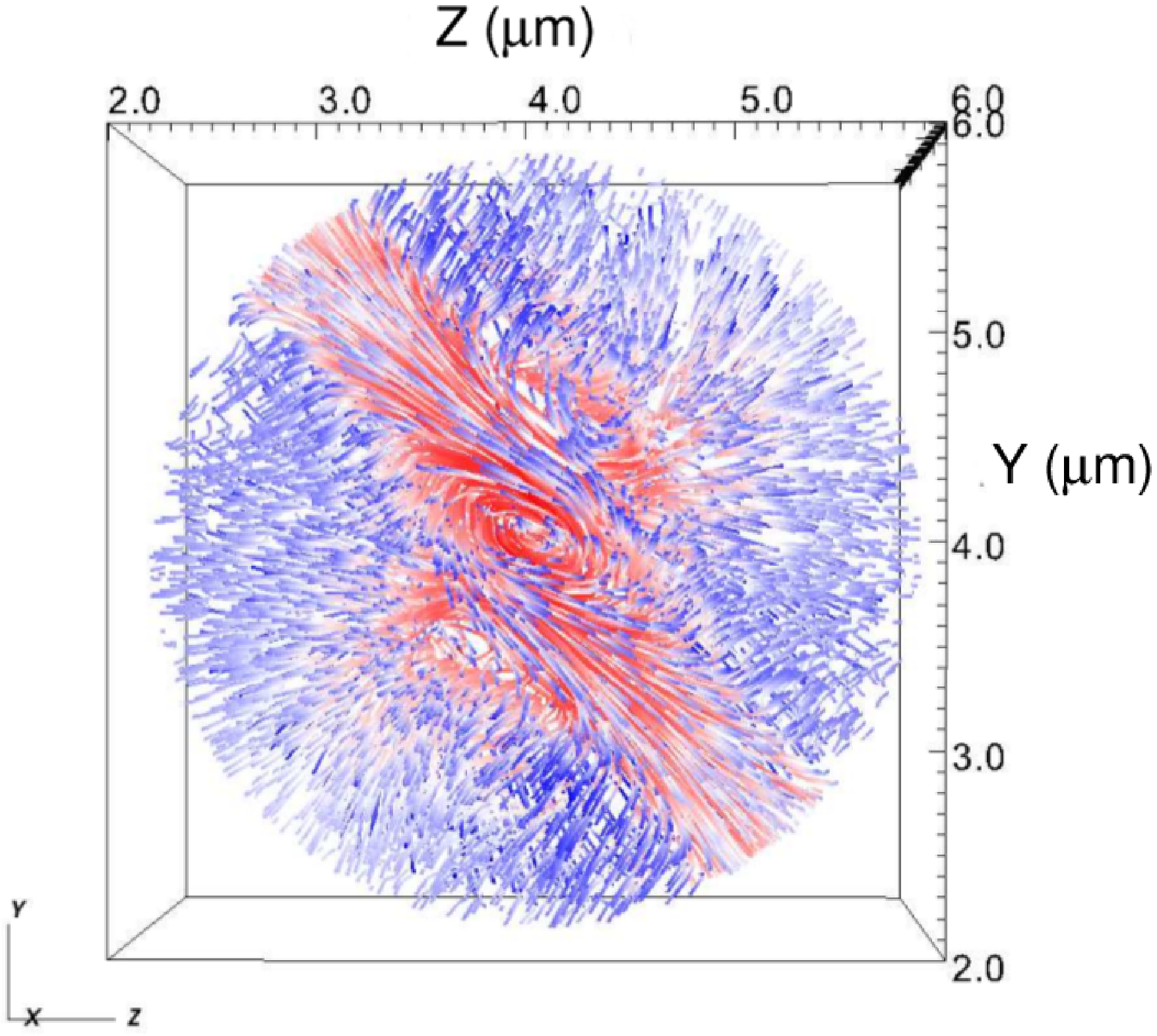}
  \caption{}
\end{subfigure}
\caption{ (a) Three-dimensional view of the electron streamlines. On the faces of the visualization box the cross section of axial magnetic field and charge density of electrons are shown. The transparent green isosurface illustrates the regions where the electric current is the highest. (b) Zoom on the streamlines in the back of the bubble. The color code is the same as in the left picture. }
\label{3Dview_stream}
\end{figure}

The complex dynamics of electrons fluxes and mechanism of the field generation is more elucidated in Fig. \ref{3Dview_stream}a, where the electron streamlines are shown flowing around the bubble in three dimensions. On the lateral sides of the simulation box the cross sections of the longitudinal magnetic field are shown, which are taken along the axis of the propagation. The back side presents the transversal cross section of the electron density at the tail of the bubble. In the right picture of Fig. \ref{3Dview_stream}b a more detailed structure of the electric currents is shown. Here one sees that a part of the electron tracks simply pass through and leave the highly concentrated region while some electrons will be captured by the electromagnetic (EM) potential and start to move on a spiral path. A small portion of the electrons moves outward and will spiral in the opposite direction (the negative B-field seen in the $xy$ plane of the left picture). 

The electron trajectory and bubble shape strongly depends on the balance between the kinetic energy of electrons gained in the force field of the laser pulse and on the EM potential generated by the bubble where the electrons are evacuated from. If the latter is high enough, electrons can be captured and accelerated while moving in the strong magnetic field. This electron dynamics can be beneficial for improving an accelerated electron beam emittance in laser plasma accelerators. The internal structure of the longitudinal magnetic field depends also on the number of captured electrons and for example if the bubble size is large or plasma density is high while the laser parameters are not adjusted correctly, the efficiency of B-field generation decreases.

\begin{figure}[h]
\centering
\hspace{6mm}
 $n_e/n_0$ \hspace{60mm}
 $B_x (kT)$ 
 
 \textbf{(a)}
\includegraphics[width=33mm]{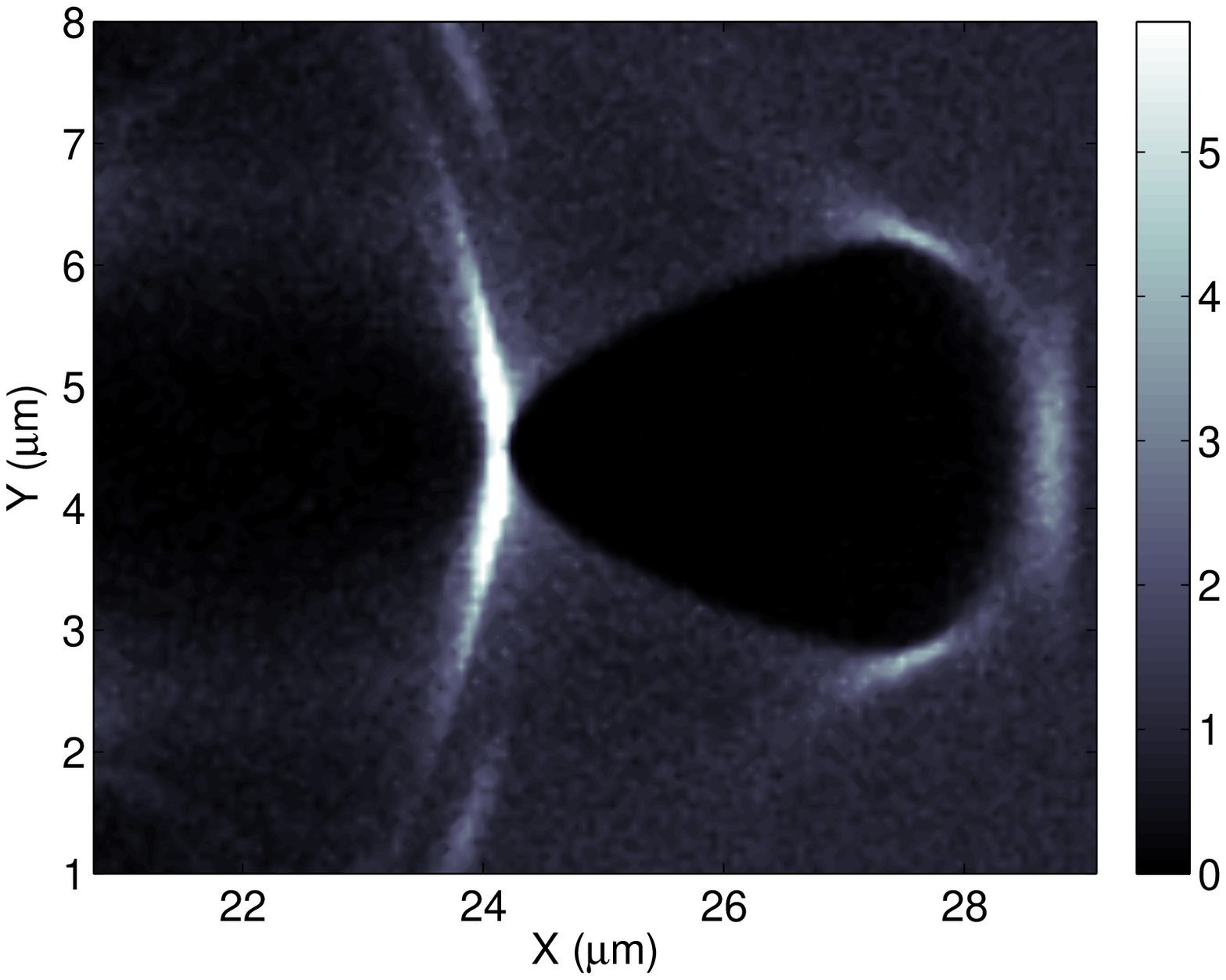}
\includegraphics[width=33mm]{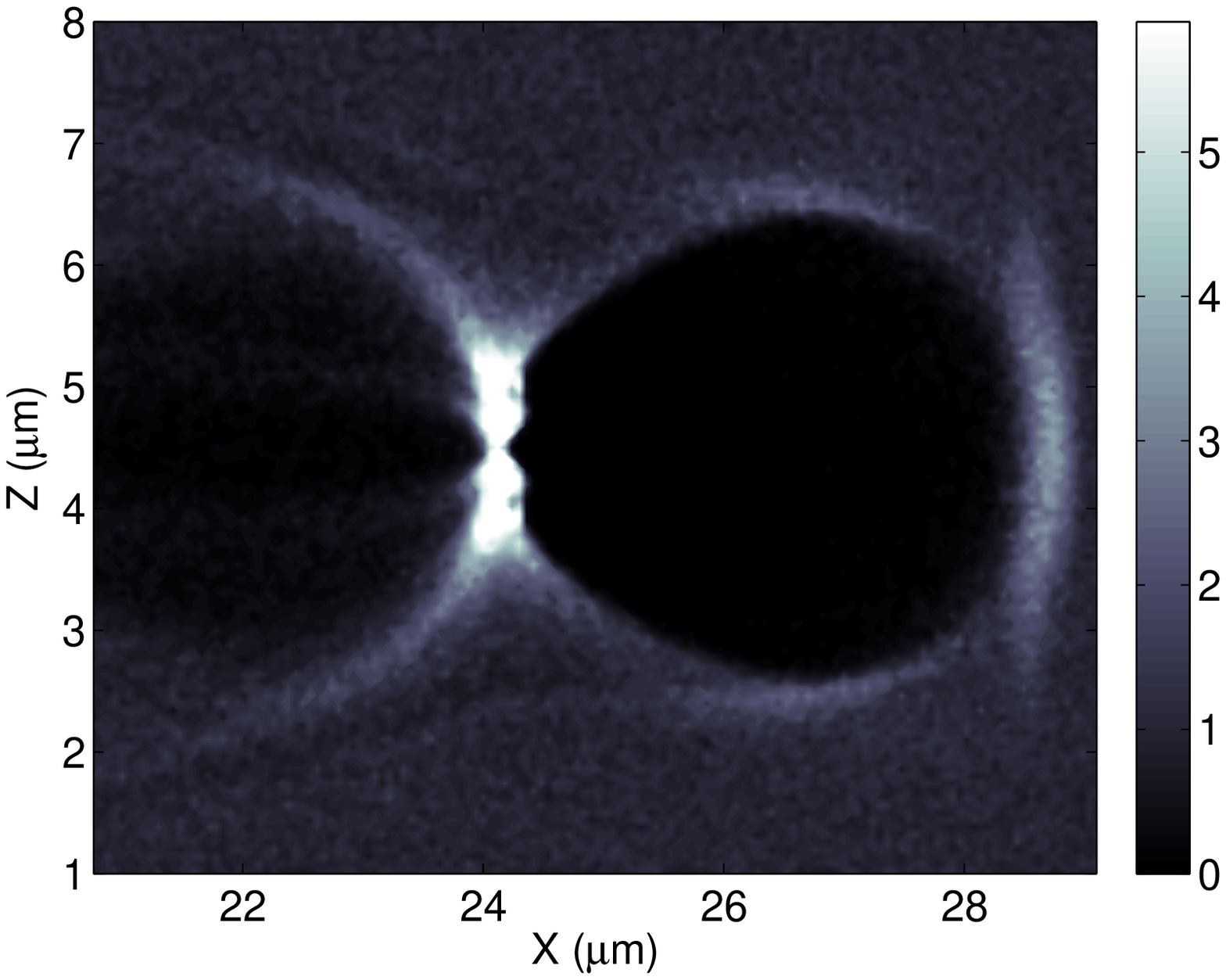}
\includegraphics[width=33mm]{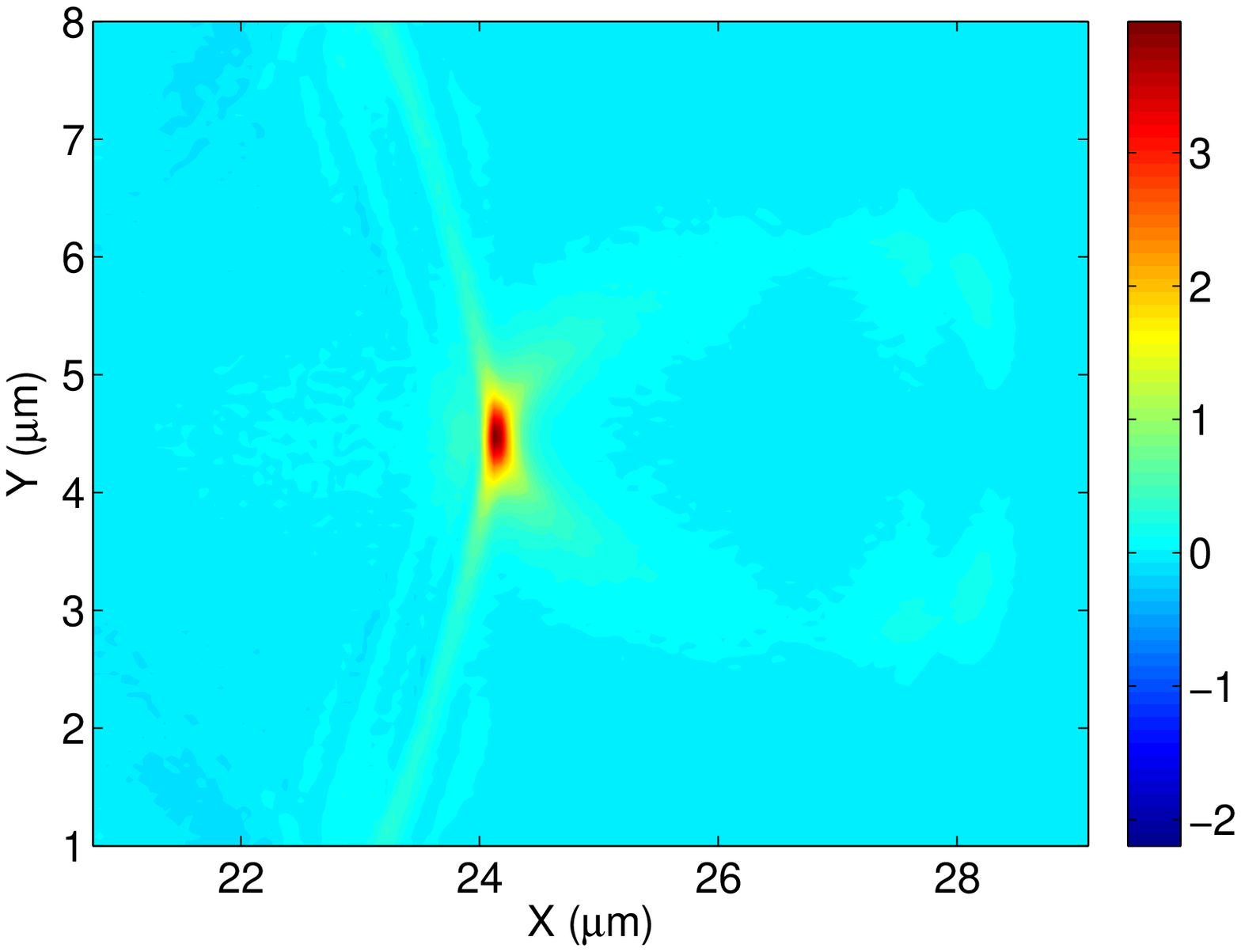}
\includegraphics[width=33mm]{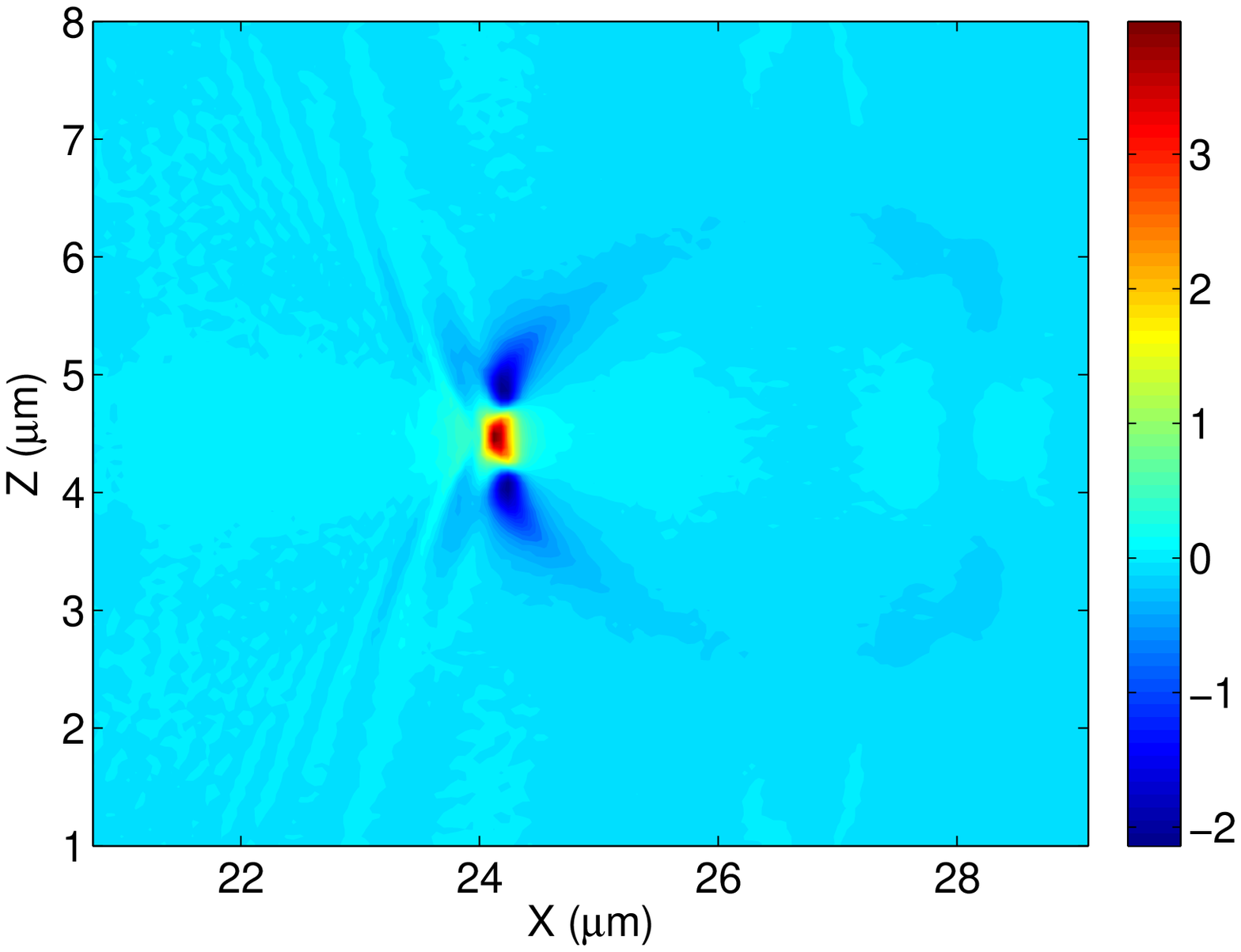}

\textbf{(b)}
\includegraphics[width=33mm]{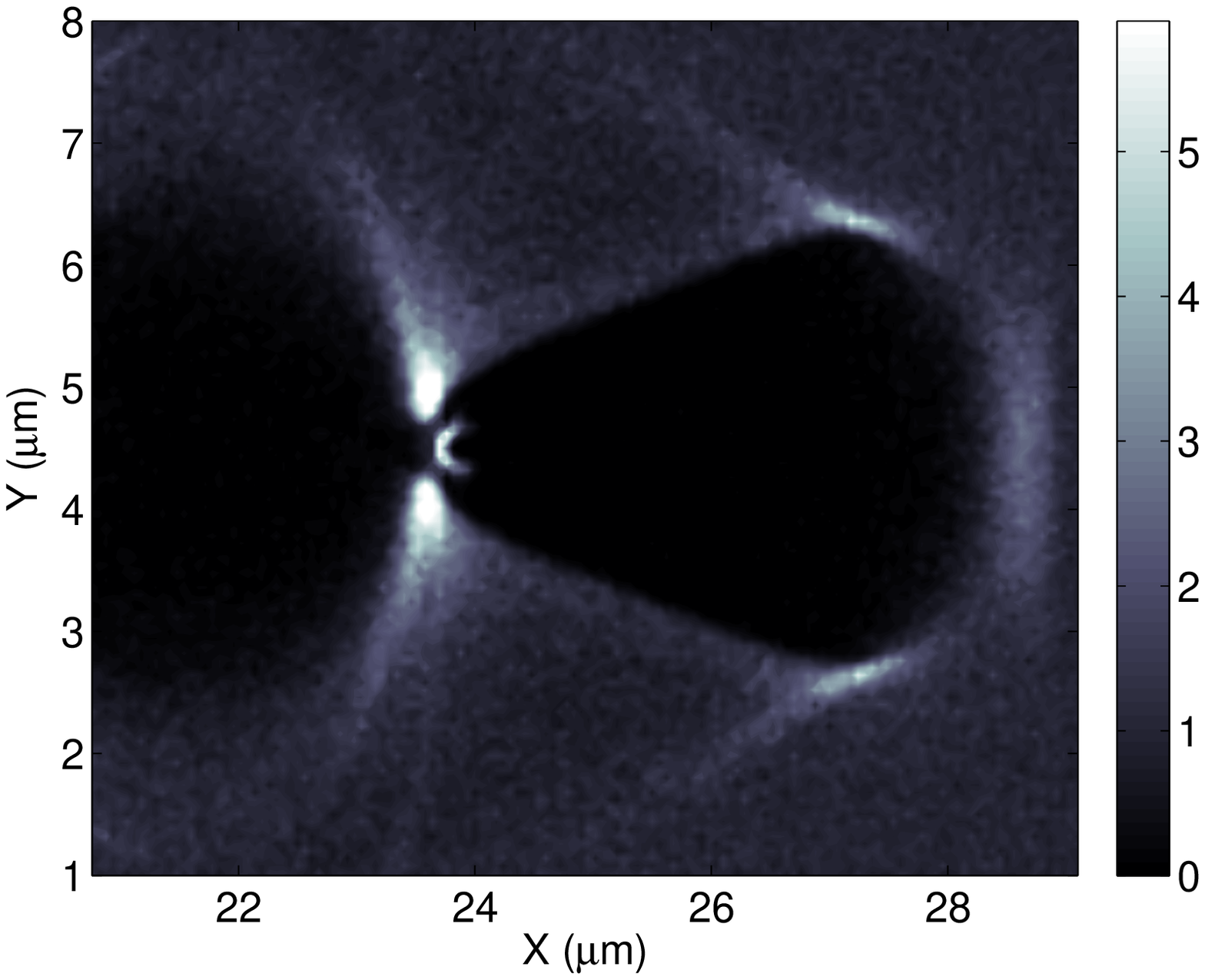}
\includegraphics[width=33mm]{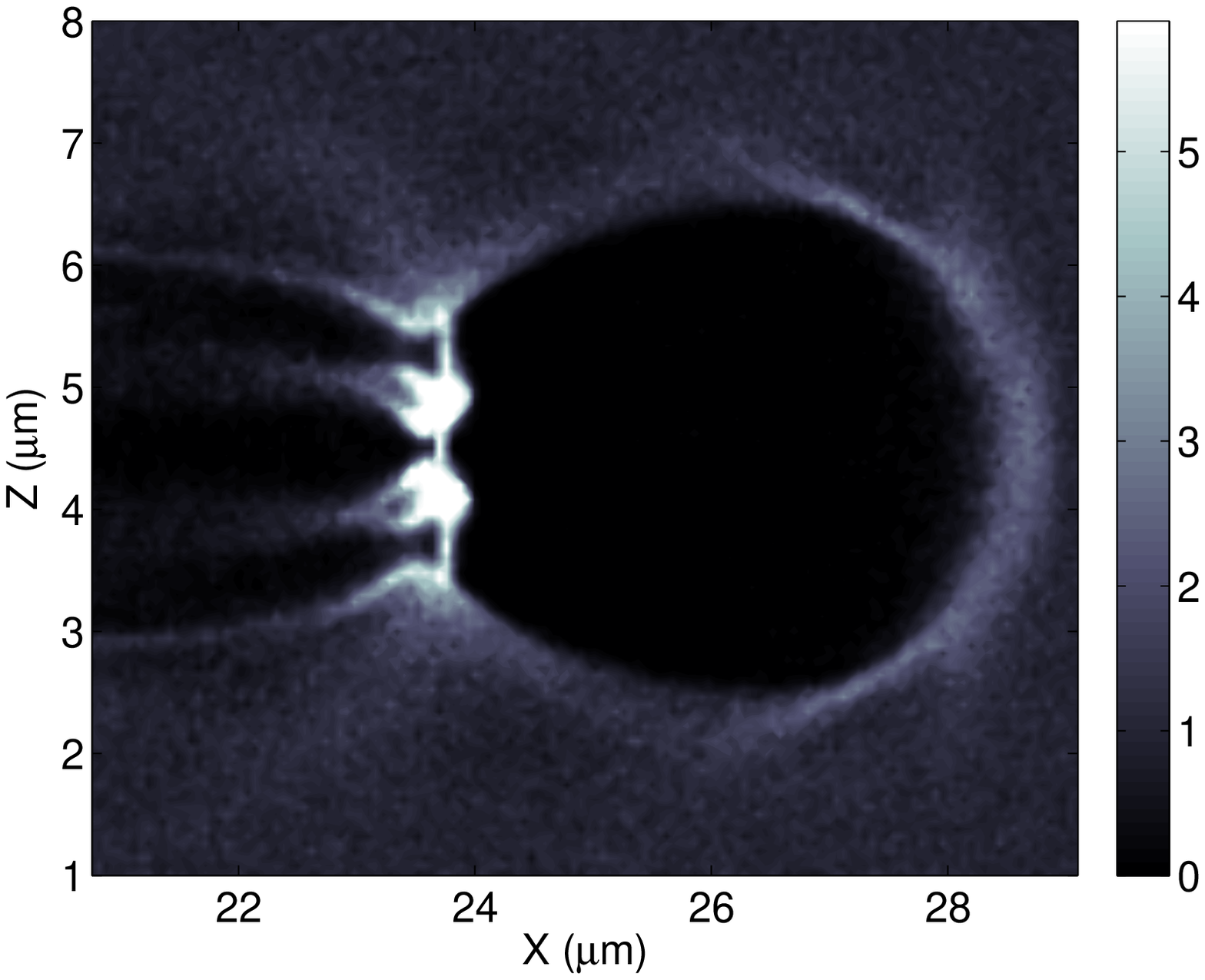}
\includegraphics[width=33mm]{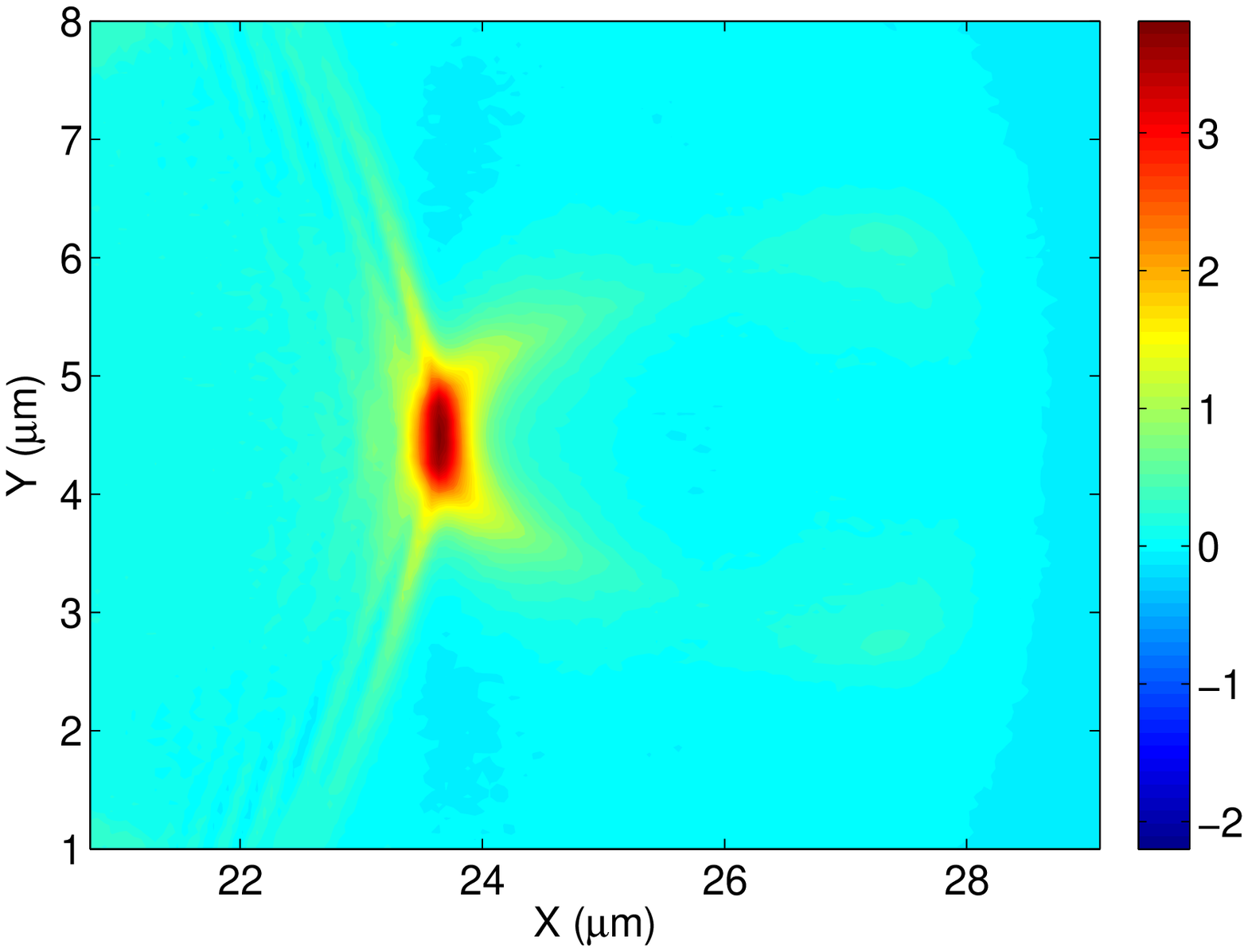}
\includegraphics[width=33mm]{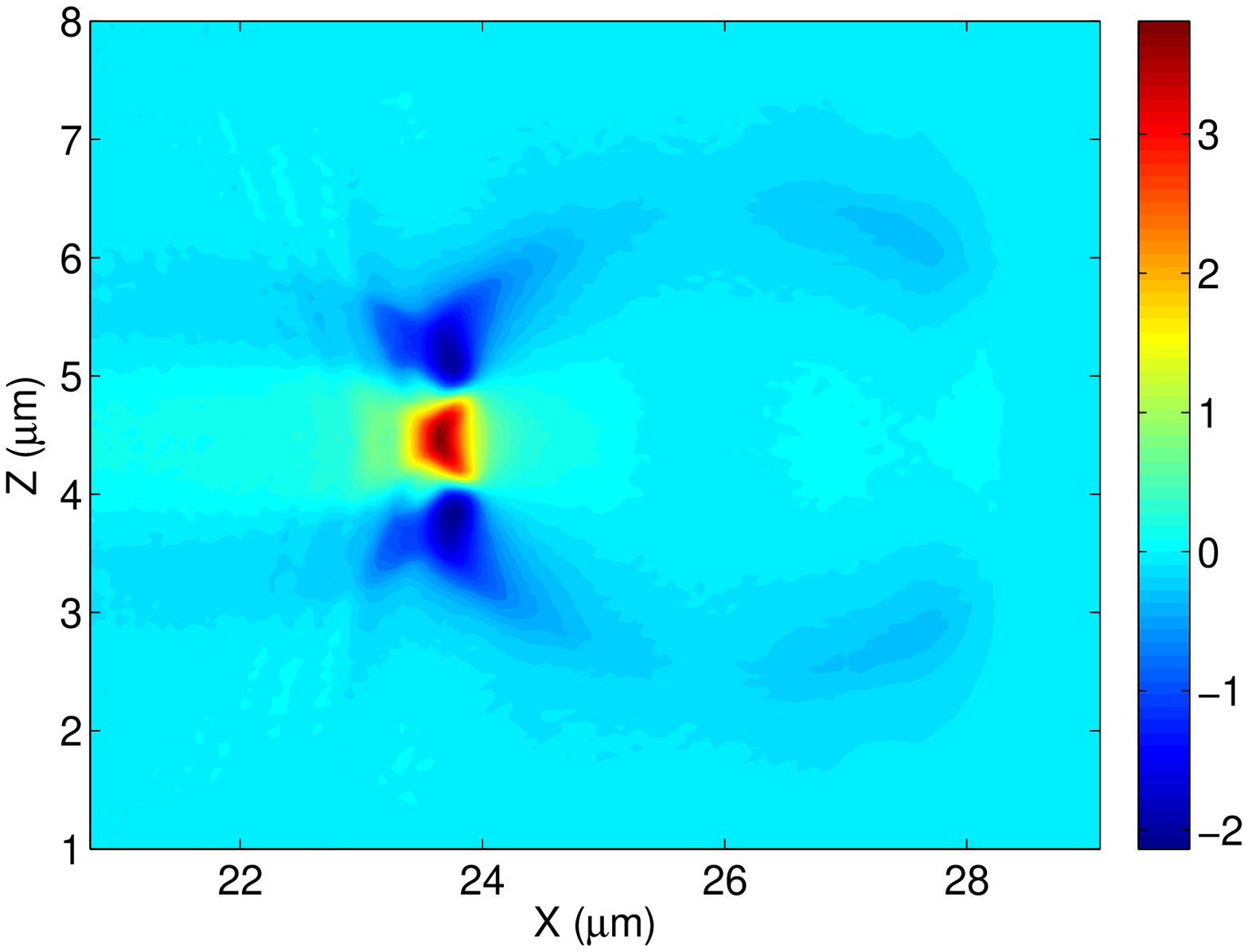}

\textbf{(c)}
\includegraphics[width=33mm]{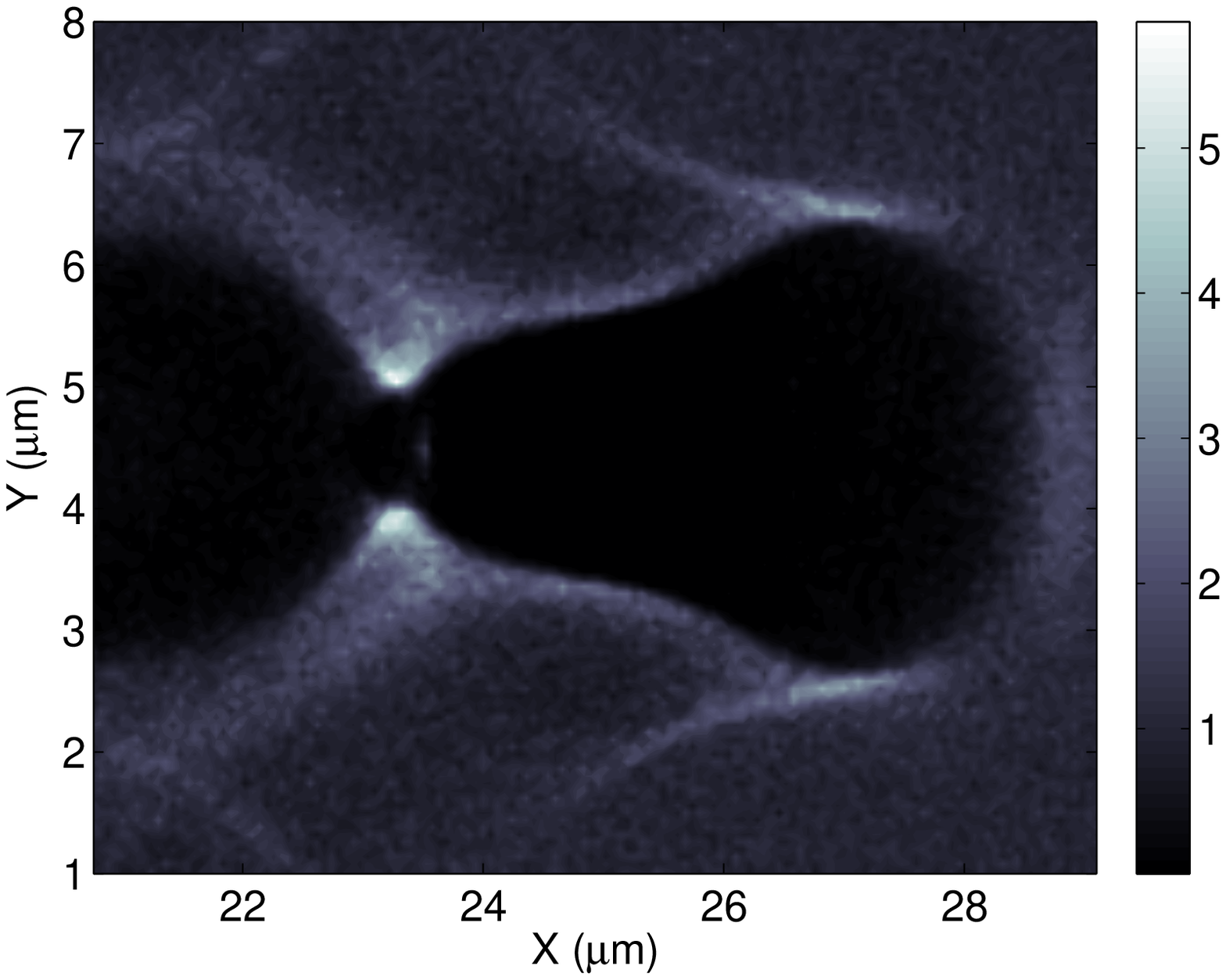}
\includegraphics[width=33mm]{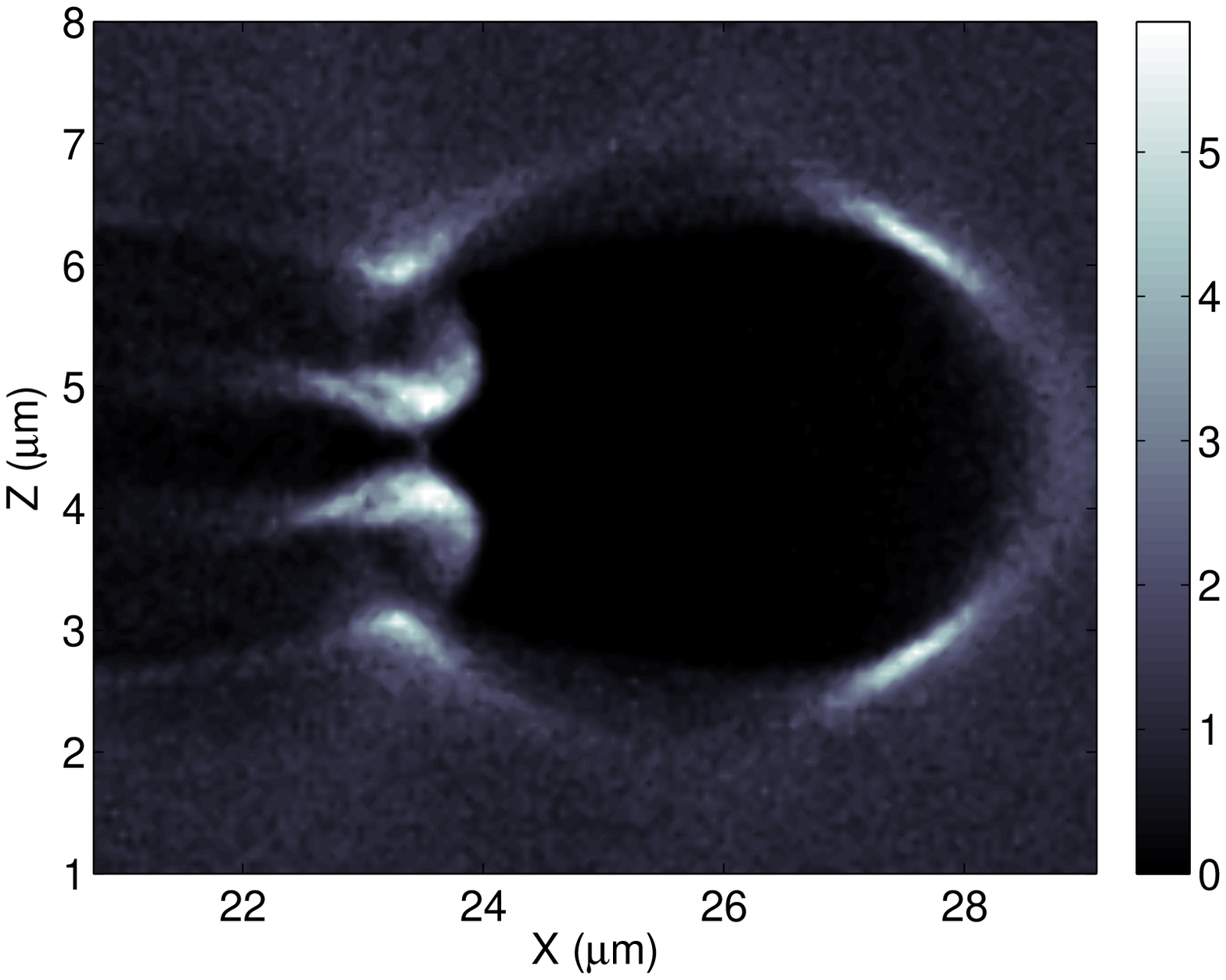}
\includegraphics[width=33mm]{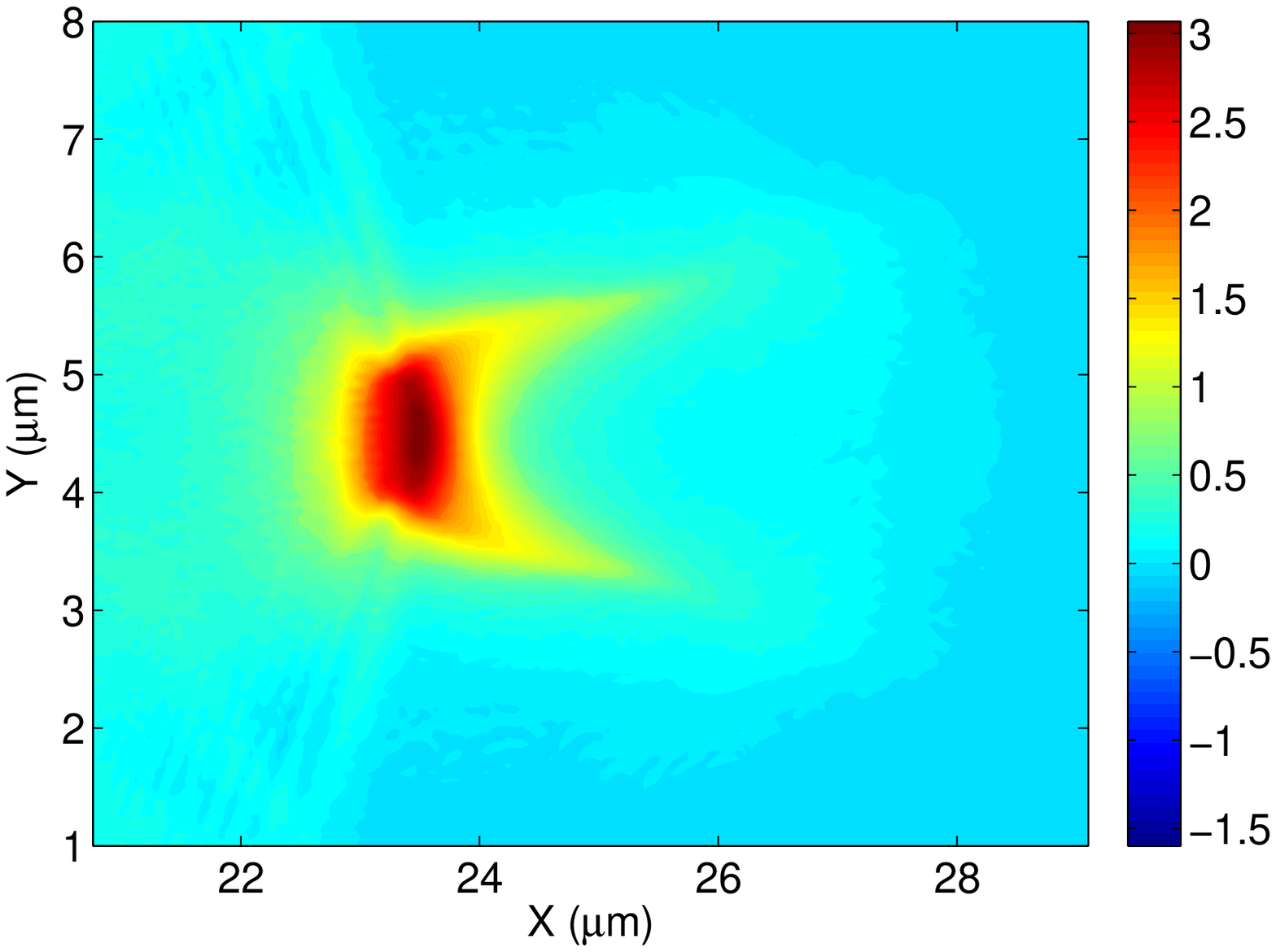}
\includegraphics[width=33mm]{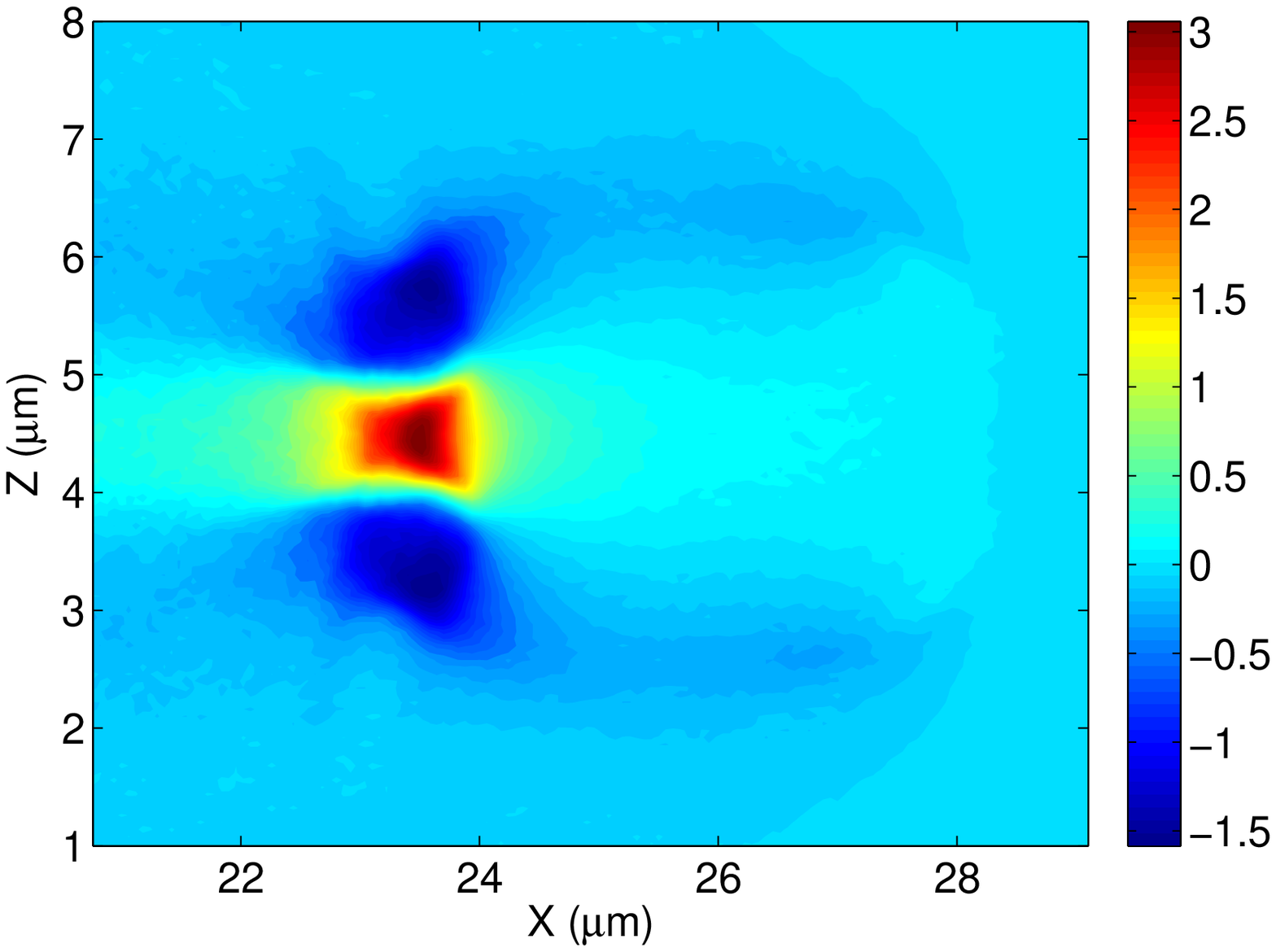}

\caption{ Cross sections of electron density normalized to initial density in the two orthogonal transversal planes and the corresponding magnetic fields in the same planes for rows: $a)$ $\lambda_{sp}=0.9 \mu$m; $b)$ $1.8 \mu$m; and $c)$ $3.6 \mu$m. In all cases $n_0=0.62\times 10^{-3}n_{cr}$.}
\label{Lsp3}
\end{figure}

\begin{figure}[h]
\centering
\textbf{(a)}\hspace{40mm}
\textbf{(b)}\hspace{40mm}
\textbf{(c)}

\includegraphics[width=47mm]{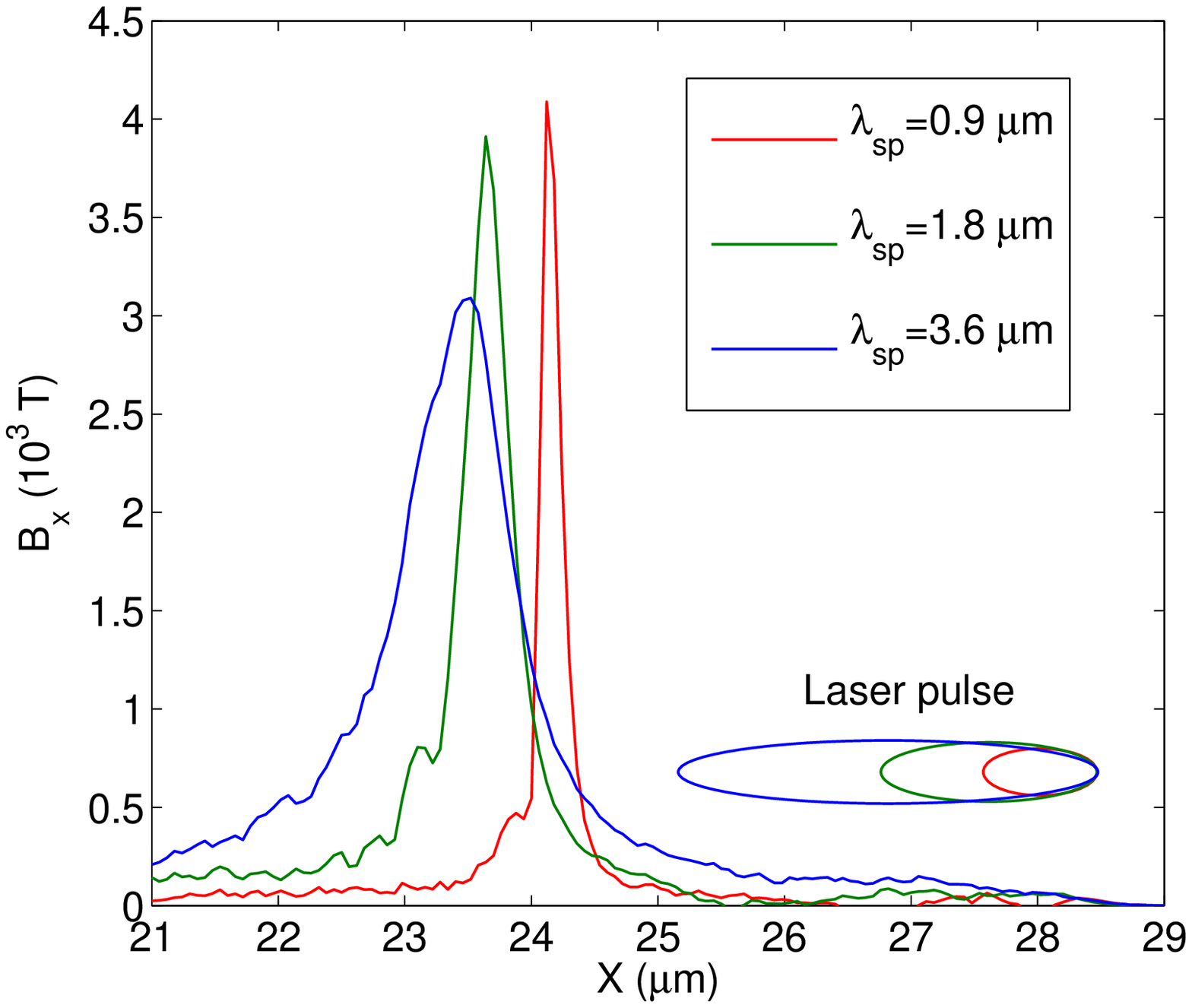}
\includegraphics[width=47mm]{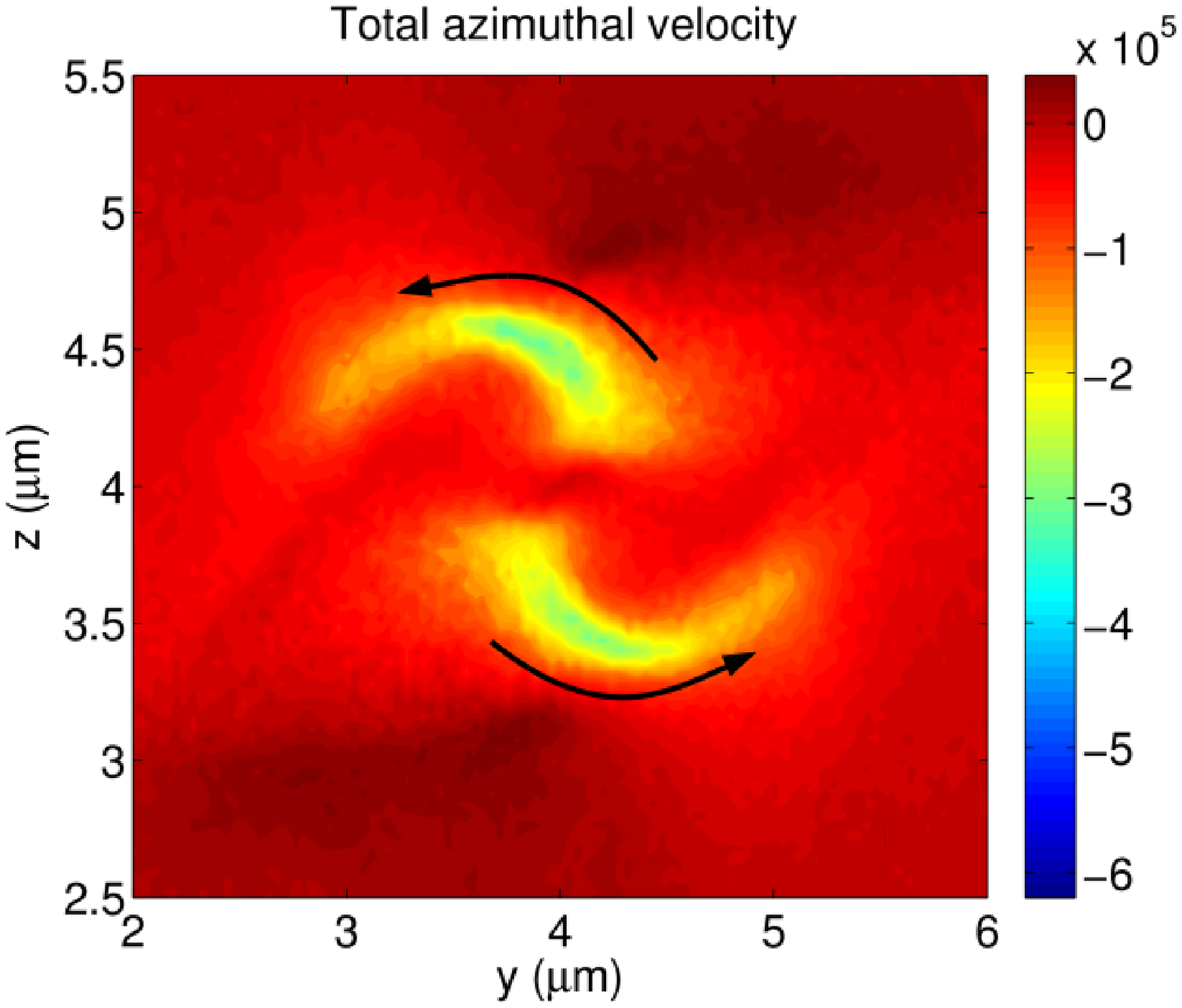}
\includegraphics[width=47mm]{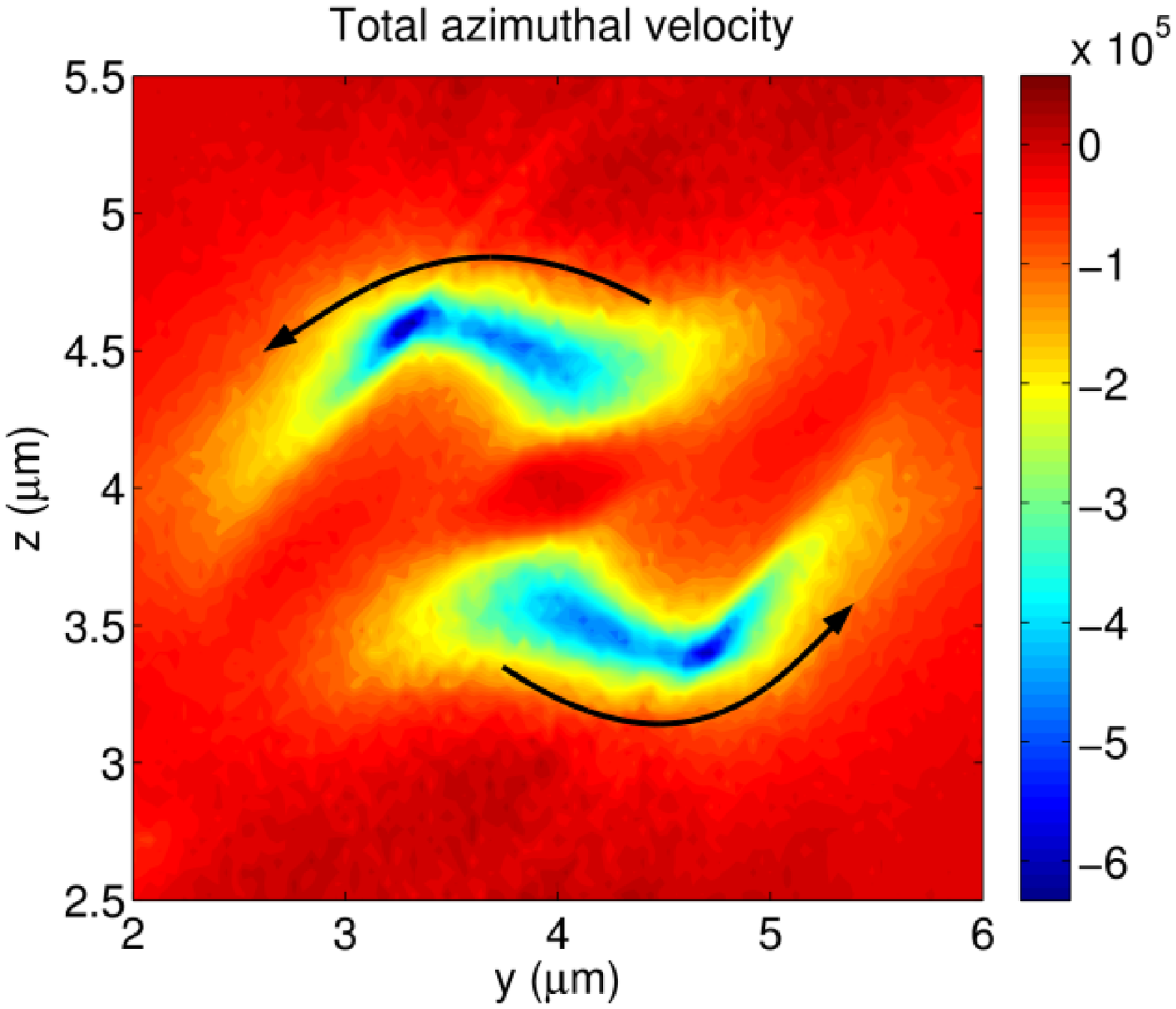}

\caption{ a) Longitudinal magnetic field along the axis of propagation for different spiral steps of the laser pulse presented in Fig. \ref{Lsp3}. b) and c) show the total azimuthal velocity ($\sum{v_\varphi/c}$) distribution in the transverse plane at the back of the bubble for  $\lambda_{sp}=0.9 \mu$m and  $\lambda_{sp}=3.6 \mu$m, respectively. The negative velocity (blue colors) indicates clockwise rotation while small positive or near zero velocities (red colors) indicates anti-clock rotation.The circle indicates the area of high compression where the highest magnetic field amplitude is achieved  }
\label{Bx_prof1}
\end{figure}

\section{Magnetic field structure and scaling with parameters}

\subsection{The bubble solenoid}

Let us consider the case when the majority of the collapsing at the end of the bubble electrons can escape the attracting field of the bubble and do not disturb the generated magnetic field. Here we will consider the dependance of the magnetic field on following set of laser and plasma parameters: plasma density, laser pulse spiral step, intensity, and wavelength. All these parameters define the electron currents outside the bubble which drive the GGs level magnetic field. In order to check the dependence on the spiral step the following simulations were performed, with the results presented in Fig. \ref{Lsp3}. In all cases the pulse length is equal to the half of the pulse spiral step, which means 180 degrees rotation of the intensity profile, but due to the Gaussian longitudinal profile the effective rotation is roughly 90 degrees (see Fig. \ref{isosurf_elvec}). It can be clearly seen that as the pulse length becomes comparable with the bubble size the magnetic field profile gets more similar to a solenoid in the $xz$ plane. The peak axial magnetic field is more localized in the case of short $\lambda_{sp}$, while it is more elongated in the case of larger $\lambda_{sp}$. In the left panels of Fig. \ref{Lsp3} the normalized electron density distribution is shown. It is clear that the charge densities in $xy$ and $xz$ planes are different as a result of "asymmetric" laser pulse. Higher compression is visible for the smaller spiral step, consequently a higher peak of magnetic field was observed, which is shown also in Fig. \ref{Bx_prof1}.

\begin{table}[h]
\begin{center}
\begin{tabular}{ |c|c|c|c|c|c|c| }
\hline

 & $2\sigma_1$ ($\mu$m) & $\lambda_{sp}$ ($\mu$m) & $\lambda_L$ (nm) & $I_0$ ( W/cm$^2$) &  $n_0/n_{cr}$ & max($B_x$) (T)  \\ \hline

Sim1 & 1.8 & $0.9 \mu$m & 100  & $1.6\times 10^{21}$ &  0.62$\times 10^{-3}$ &  4000  \\  \hline
Sim2 & 1.2 & $0.9 \mu$m &  100  & 8$\times 10^{21}$ & 6.2$\times 10^{-3}$ & $2.8\times 10^4$ \\ \hline

Sim3 & 0.8 & $0.9 \mu$m & 100  & 3.2$\times 10^{22}$ &  0.062 & $2.5\times 10^5$  \\ \hline
Sim4 & 0.3 & $0.3 \mu$m & 20  & 8$\times 10^{23}$ &  0.025 &   $0.95\times 10^6$ \\ \hline

Sim5 & 14.4 & $7.2 \mu$m &  800  & $1.6\times 10^{21}$ &  0.04 &  $1.2\times 10^4$  \\  \hline

Sim6  & 14.4 & $7.2 \mu$m &  800  & $0.8\times 10^{21}$ &  0.01 &  5000  \\  \hline

Sim7  & 14.4 & $7.2 \mu$m &  800  & $3.2\times 10^{21}$ &  0.01 &  8000 \\  \hline

Sim8 & 9.6 & $7.2 \mu$m & 800  & 2$\times 10^{22}$ & 0.1 &  $5\times 10^4$ \\ \hline

\end{tabular}
\end{center}
\caption{ Simulation parameters. In all simulations $\sigma_2=\sigma_1/2$. }
\label{simParam}
\end{table}

\begin{figure}[h]
\centering

\textbf{(a)}\hspace{40mm}
\textbf{(b)}

\includegraphics[width=59mm]{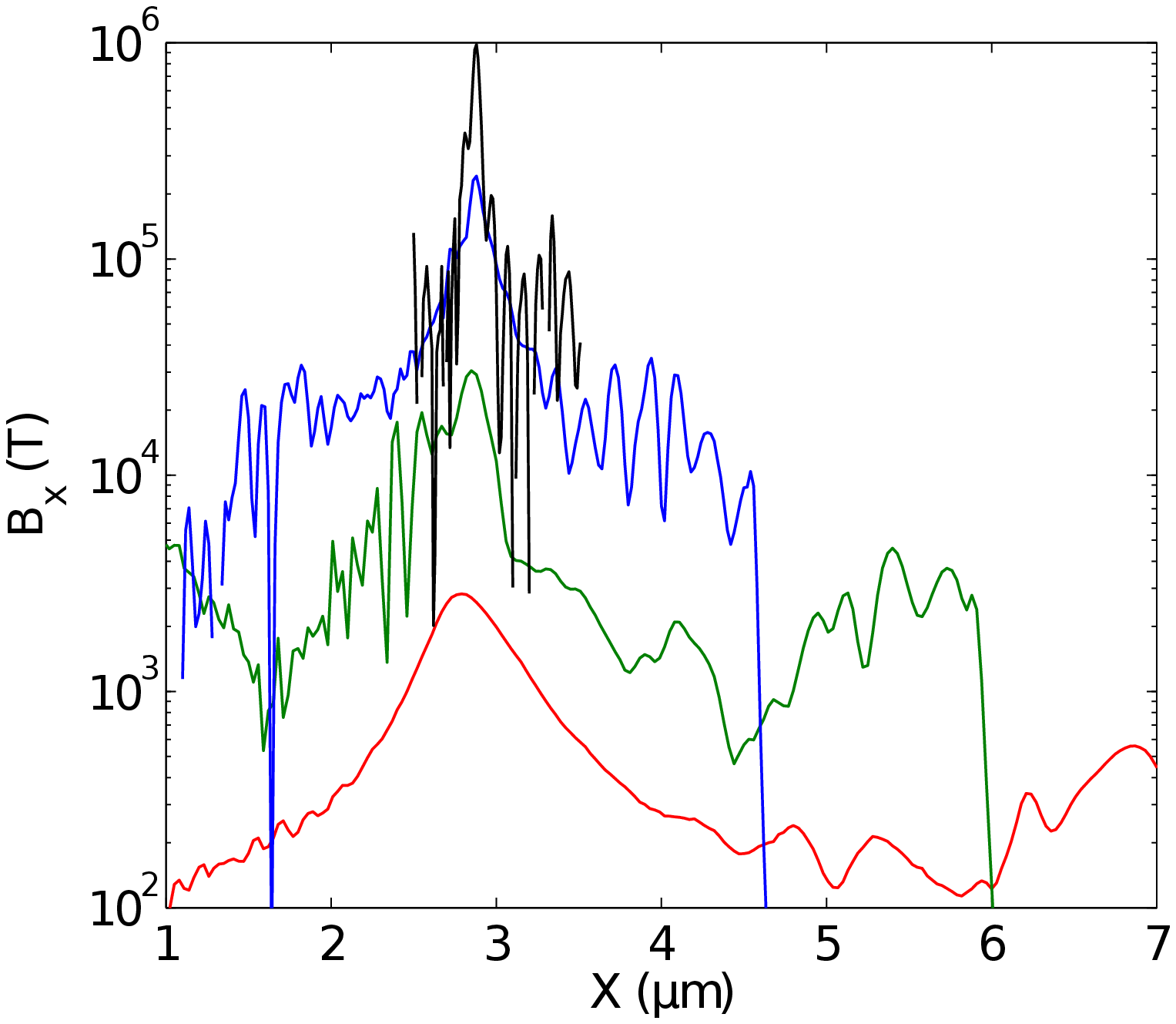}
\includegraphics[width=55mm]{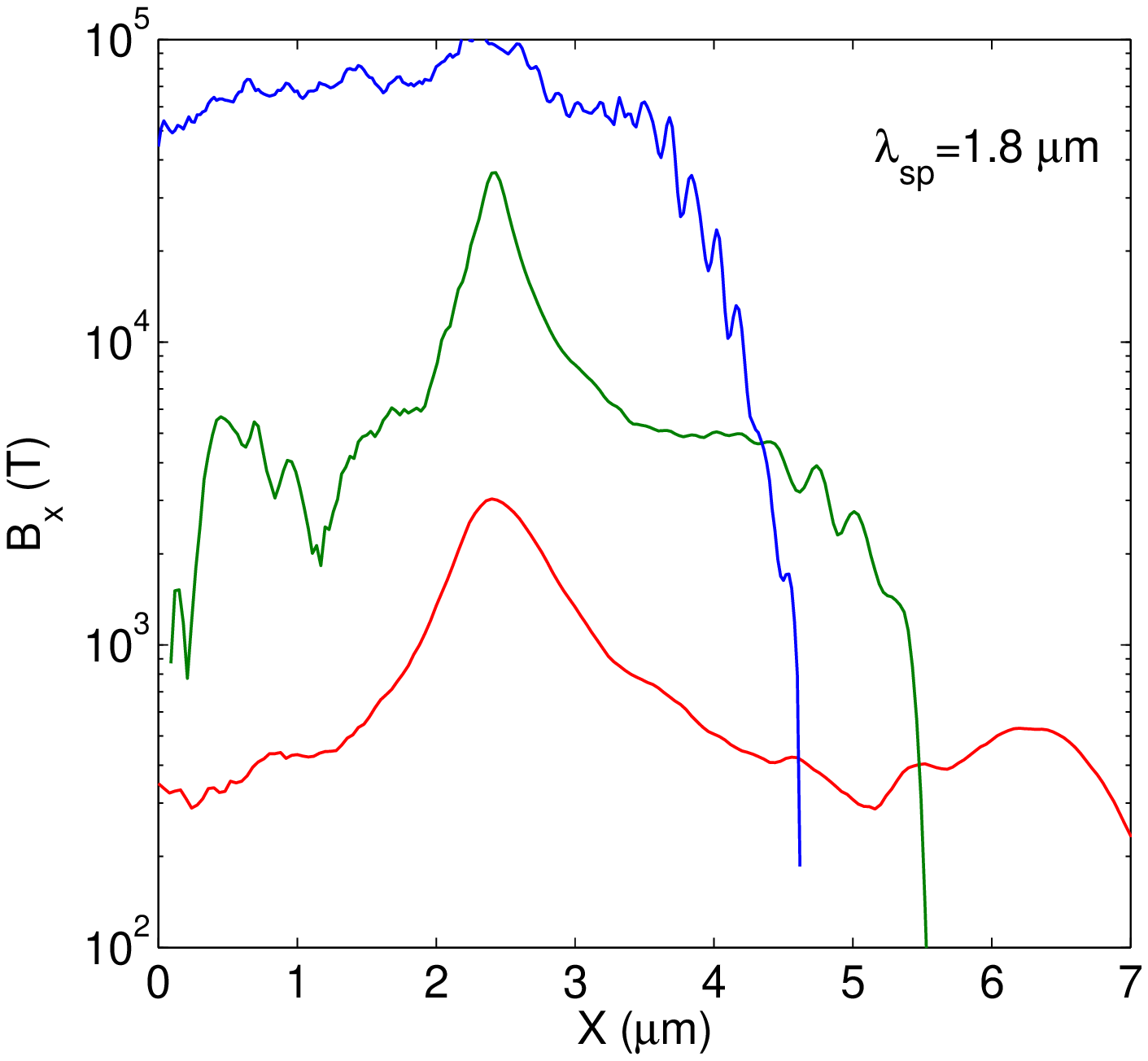}
\caption{  Longitudinal magnetic field along the axis of propagation for parameters shown in: (a) Table \ref{simParam} and (b) for two times larger spiral step. The red, green, blue and black lines corresponds to simulations $1,2,3,4$ respectively. The  color code is the same in both pictures.  }
\label{Bx_prof2}
\end{figure}

Indeed the shape of the magnetic field distribution and the actual peak value depend on the details of complex electron's trajectories and on the compression level which they acquire in the tail of the bubble. However the self induced repulsive forces can result in destruction of the currents and thus limiting the amplitude of generated magnetic field. If we assume that the transversal size of the bubble does not change (which is true in Fig. \ref{Lsp3}) the compressed electron density scales as $n_e=(\lambda_p/\lambda_{sp})n_0$. The second thing to be noticed is the fact that the magnetic field is generated by finite current sheets with thickness approximately $\lambda_{sp}/2$. Applying Ampere's law to this geometry one finds for the magnetic field $B=\mu_0 j_0 \lambda_{sp}$, where $j_0=en_ev_{\varphi}$ and $v_{\varphi}$ is the azimuthal velocity. Inserting the expression for the compressed electron density and considering that some of the evacuated electrons contribute to the transversal current we find $B= p\mu_0 e n_0 \lambda_p v_{\varphi}/2$, where $p<1$ indicates the percentage of electrons contributing to the transverse current.  If one uses $\lambda_p\approx 8 \mu$m the estimated value of the B-field is $B \approx p\cdot 16$ kT, which is close to the measured one if assume that only quarter of the electrons are contributing to the transverse current.

According to Fig. \ref{Bx_prof1}a the peak magnetic field depends only weakly on the spiral step (or pulse length in our setup), but through the expression of ponderomotive force we find that

\begin{equation}\label{eq:gamma}
\gamma \sim \nabla I_L \lambda_L^2 \sim I_0\lambda_L/l,
\end{equation}
where $l=\lambda_{sp}/\lambda_L$. On the other hand the bubble length (or half plasma wavelength) is proportional to $\sqrt{\gamma/n_0}$, thus

\begin{equation}\label{eq:Bfield}
B\sim (\gamma n_0)^{1/2} v_{\varphi}.
\end{equation}

This scaling indicates that the axial magnetic field shown in Fig. \ref{Bx_prof1}a should be two times lower for 4 times longer spiral step. However, the peak values are much closer to each other, which can be attributed to the larger azimuthal momentum acquired with larger $\lambda_{sp}$. The comparison of azimuthal velocity distribution in the back of the bubble is shown in Fig. \ref{Bx_prof1}b and \ref{Bx_prof1}c for  $\lambda_{sp}=0.9 \mu$m and  $\lambda_{sp}=3.6 \mu$m, respectively. The number of rotating electrons is the same in both cases, but their azimuthal velocity is two times higher in the case of longer spiral step, which compensates for the reduced gamma factor. 

It is straightforward to expect higher magnetic field inside larger volumes by increasing the laser intensity (energy available to drive the current) and plasma density (current available to generate the field). In the following studies it is proven that with the laser intensity reachable with the current technology the GGs level magnetic fields can be generated in the experiments in the laboratory environment. The parameters used in the next simulations are shown in Table \ref{simParam}. The first parameter-set (Sim1) has been used in the previous simulations. The bubble size is defined by three parameters: laser spot size ($W_L=2\sigma_1$), laser intensity and plasma density. In order to have similar bubble structure as before we chose the laser spot size to be between one third and half of the bubble length.

\begin{figure}[h]
\centering
\textbf{(a)}\hspace{50mm}
\textbf{(b)}

\includegraphics[width=57mm]{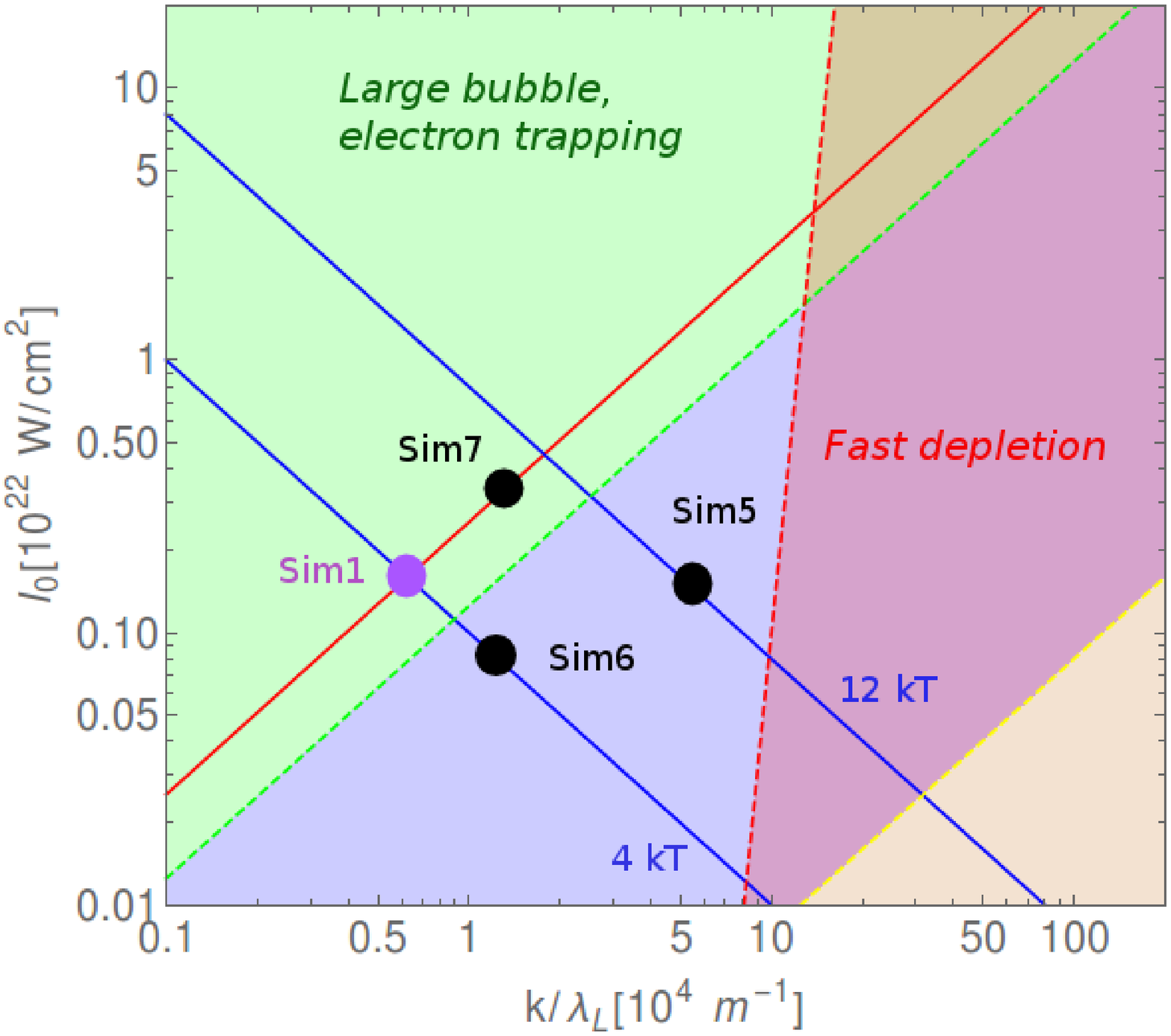}
\includegraphics[width=57mm]{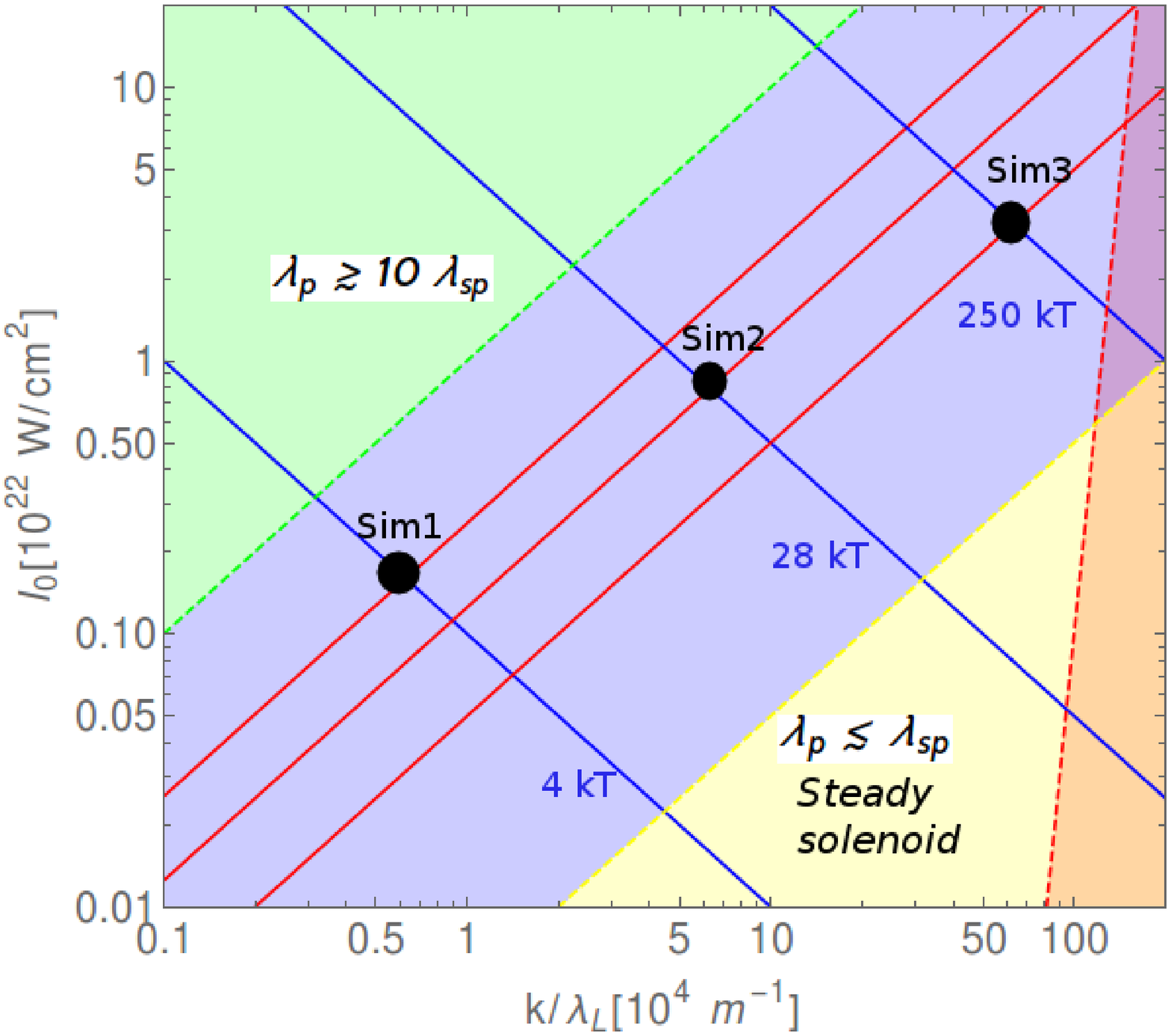}
\caption{ Parameter-map of the mechanism, where the region of different regimes are also indicated, for two laser wavelengths: a) $\lambda_L=800$ nm and b) $\lambda_L=100$ nm. The blue and red lines (Eq. (\ref{eq:cond1}, \ref{eq:cond2})) show the parameters required to generate the same magnetic field amplitude and the same field distribution, respectively. The blue area shows the parameter interval discussed in this section: the bubble solenoid regime. The red dashed line corresponds to $k=n_0/n_{cr}=0.1$, behind which the laser pulse depletion affects the magnetic field generation and the model should be revisited.  }
\label{isocurves}
\end{figure}

The axial magnetic fields measured for the first four parameter-sets are shown in Fig. \ref{Bx_prof2}a and \ref{Bx_prof2}b. The expression derived above, in the strongly relativistic case, i.e. $v_{\varphi}\approx c$, predicts a scaling $B \sim (\gamma n_0)^{1/2}$, which is close to the observed scaling. In the case of larger spiral step (Fig. \ref{Bx_prof2}b) the third simulation did not give the expected result, because the pulse was longer than the bubble and the magnetic field generation enters a different regime, discussed later. In the case of Sim4 the plasma density is near the solid density, which allows the formation of very small bubble, therefore the laser wavelength and pulse size had to be decreased. Although the resulting magnetic field reaches the MT level, the required laser parameters do not seem to be reachable at the current stage of laser technology. However, it is possible to scale the parameters such that the resulting B-field remains the same. For this the following condition has to be fulfilled: 

\begin{equation}\label{eq:cond1}
n_0\lambda_p\sim \sqrt{n_0I_0\lambda_L/l}\sim \sqrt{kI_0/(\lambda_L l)}\approx constant,
\end{equation}
where $k=n_0/n_{cr}$. In order to have the same bubble shape the following constrain can be added: 

\begin{equation}\label{eq:cond2}
\lambda_p/\lambda_{sp}\sim \sqrt{I_0/(\lambda_L l^3n_0)} \sim \sqrt{I_0\lambda_L/(l^3k)} \approx constant. 
\end{equation}
Unfortunately by increasing the laser wavelength up to 0.8 $\mu$m the plasma becomes overdense which results in quick depletion (absorption or even reflection) of the laser pulse and no axial B-field is generated. At this stage we can speculate that the issue can be resolved if for instance an electron beam with angular momentum is used to drive the currents capable of generating MT magnetic fields. 

From the expressions presented above it follows that either the bubble shape or magnetic field amplitude changes by modifying the laser or plasma parameters. In the above simulations the dimensionless pulse length was $l=9$ and if it is kept the same, the laser wavelength and intensity can be tuned, within well defined limits, to obtain the same magnetic field strength. In Fig. \ref{isocurves} the iso-value curves of max($B_x$) and $\lambda_p/\lambda_{sp}$ are shown, which indicate the parameter plane scaling for constant pulse length $l$. The simulations discussed and presented in Table \ref{simParam} are shown on these planes of parameters. The blue area indicates the parameter space where the bubble regime can be observed. The green area shows the regime of strong electron trapping (not favorable for generation of large magnetic fields), the yellow area corresponds to the regime of steady solenoid (presented in the next section) and the red area is the area where the plasma density is large and pulse energy depletion is not negligible any more. The yellow and green dashed lines are not exact curves and due to energy depletion transitions from green to blue and from red to yellow are always expected. It is clearly visible that the laser wavelength strongly influences the diagram and it pushes the blue area towards lower intensity and density regions, which in turn means weaker magnetic field generation. One notices that for the generation of high B-field in static bubble solenoid high intensity and short wavelength is required.

\begin{figure}[h!]
\centering
\textbf{(a)}\hspace{40mm}
\textbf{(b)}\hspace{40mm}
\textbf{(c)}

\includegraphics[width=45mm]{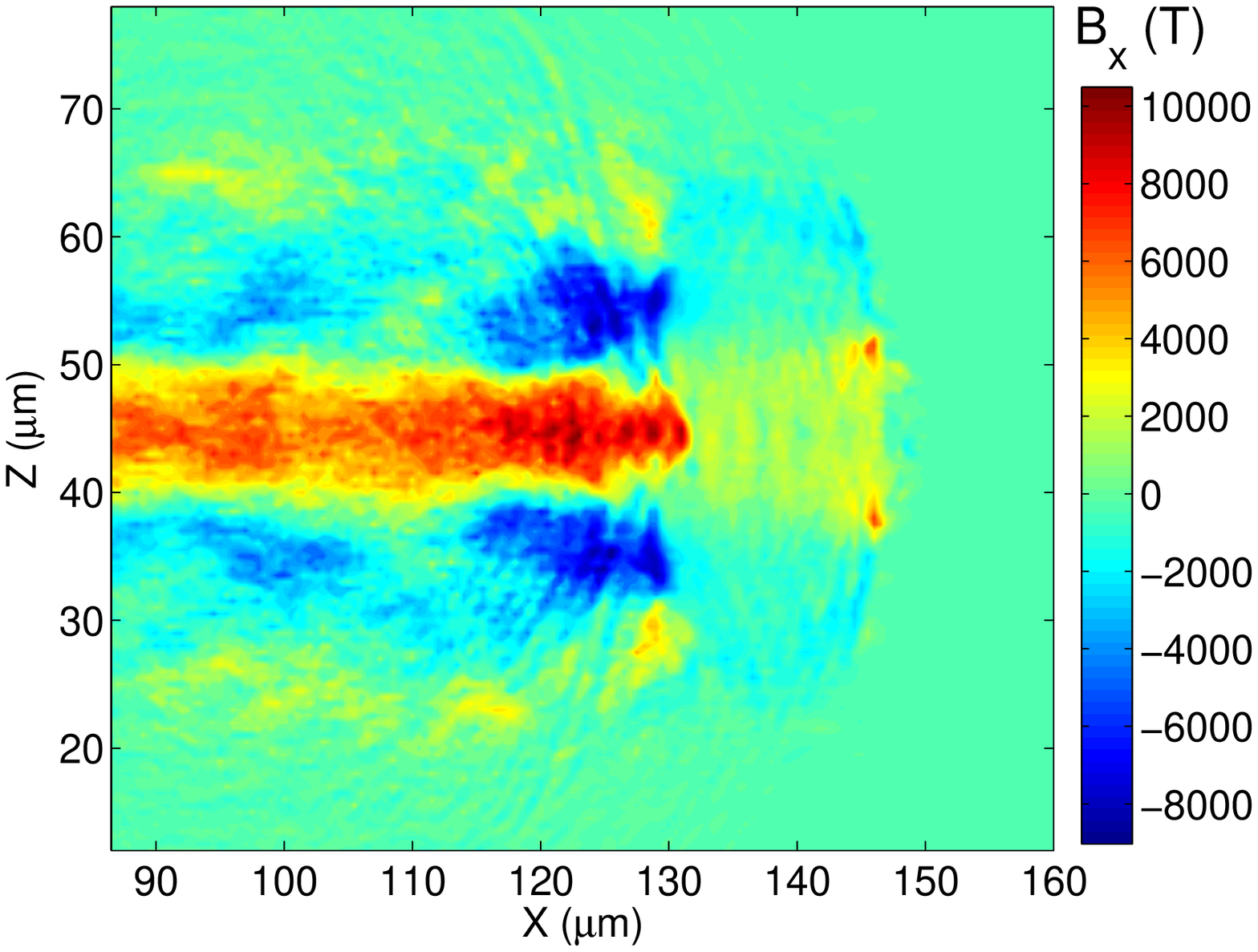}
\includegraphics[width=45mm]{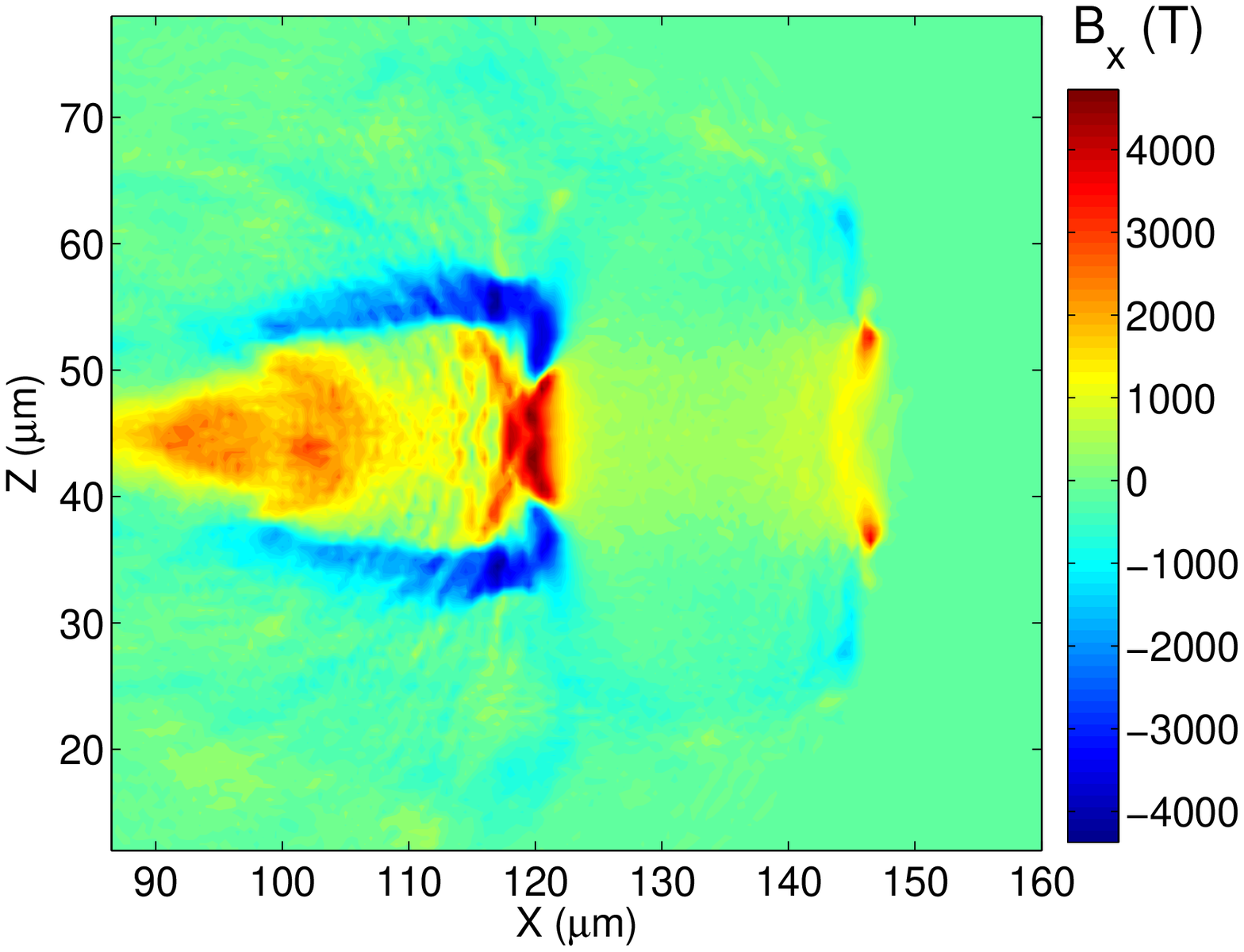}
\includegraphics[width=45mm]{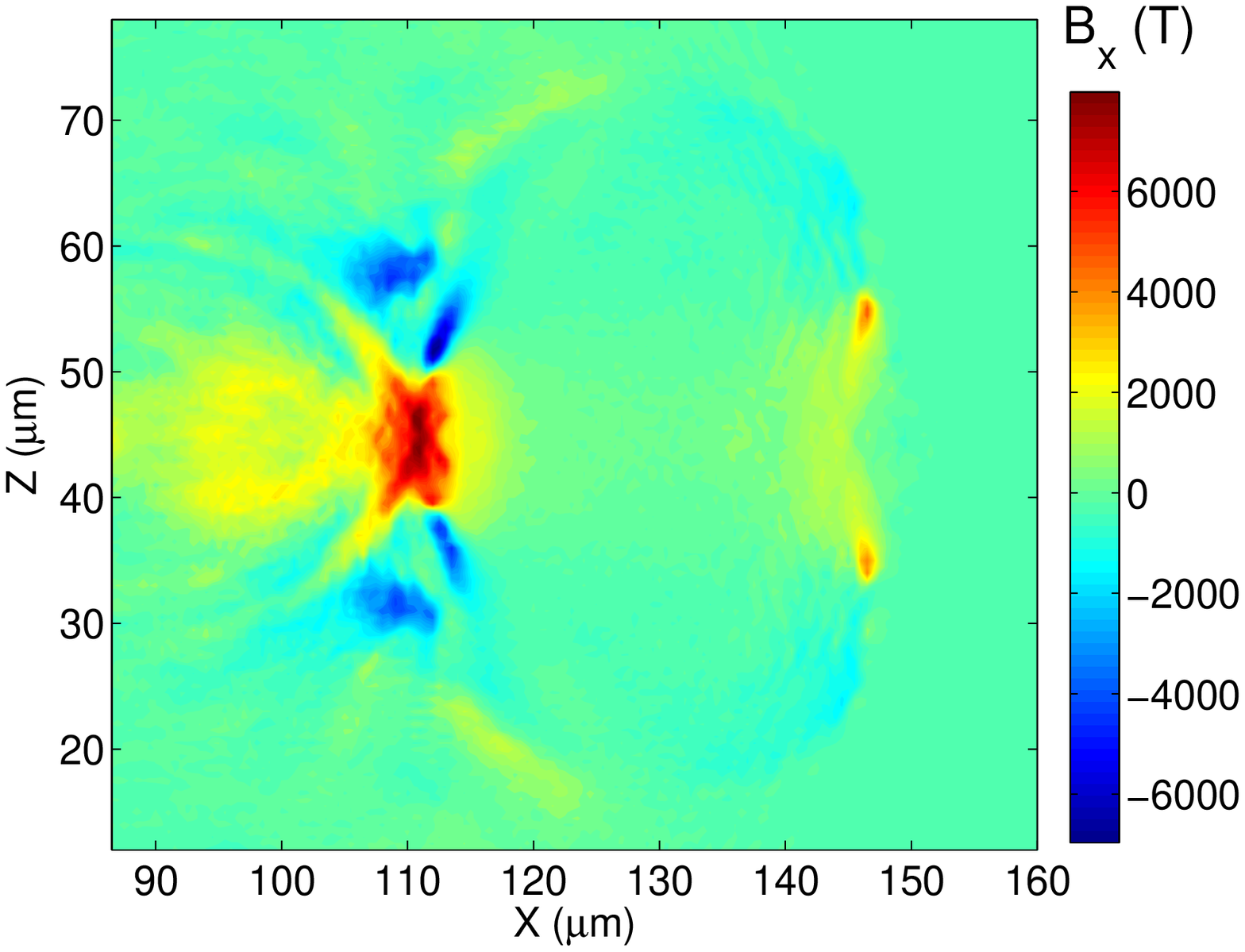}

\includegraphics[width=45mm]{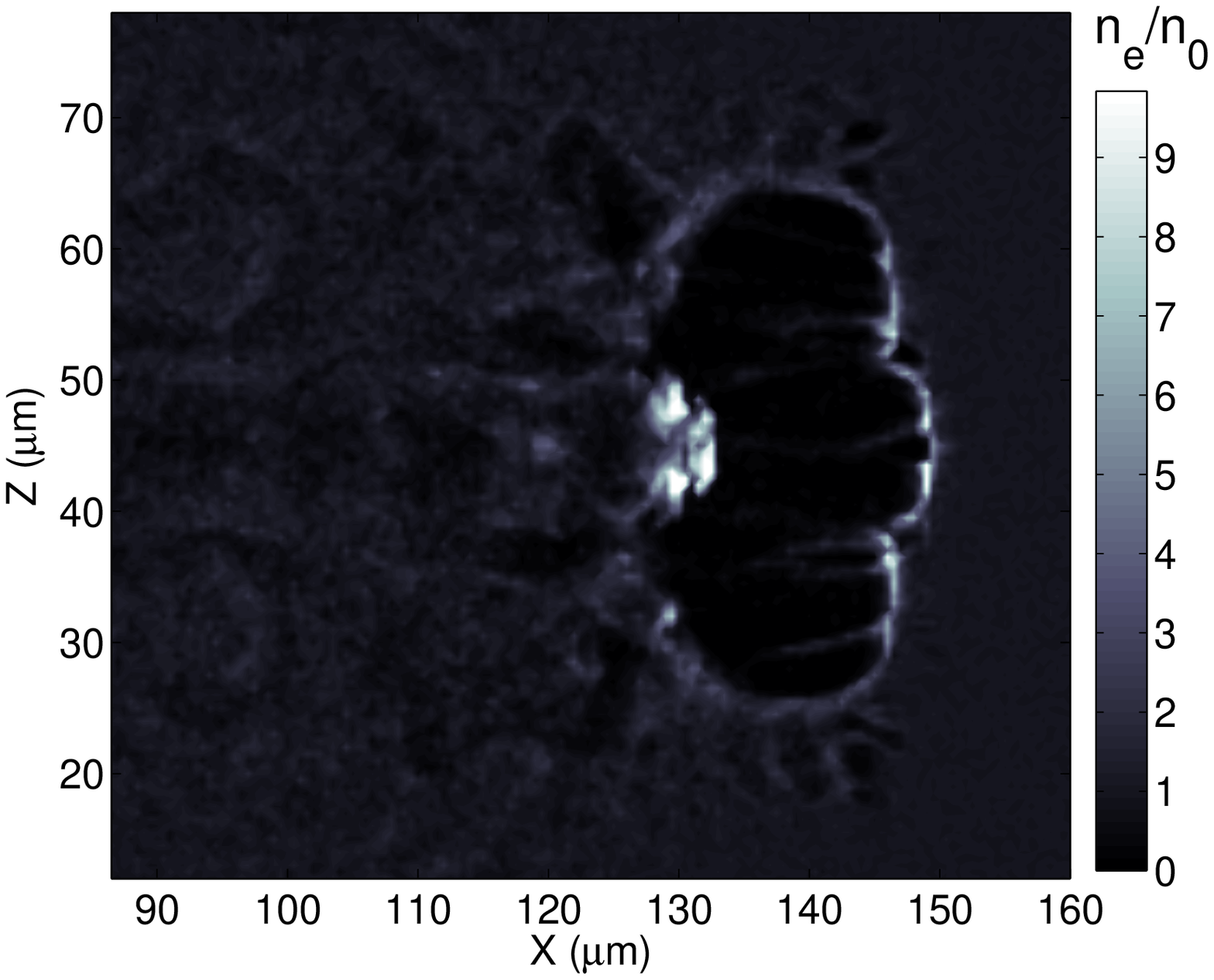}
\includegraphics[width=45mm]{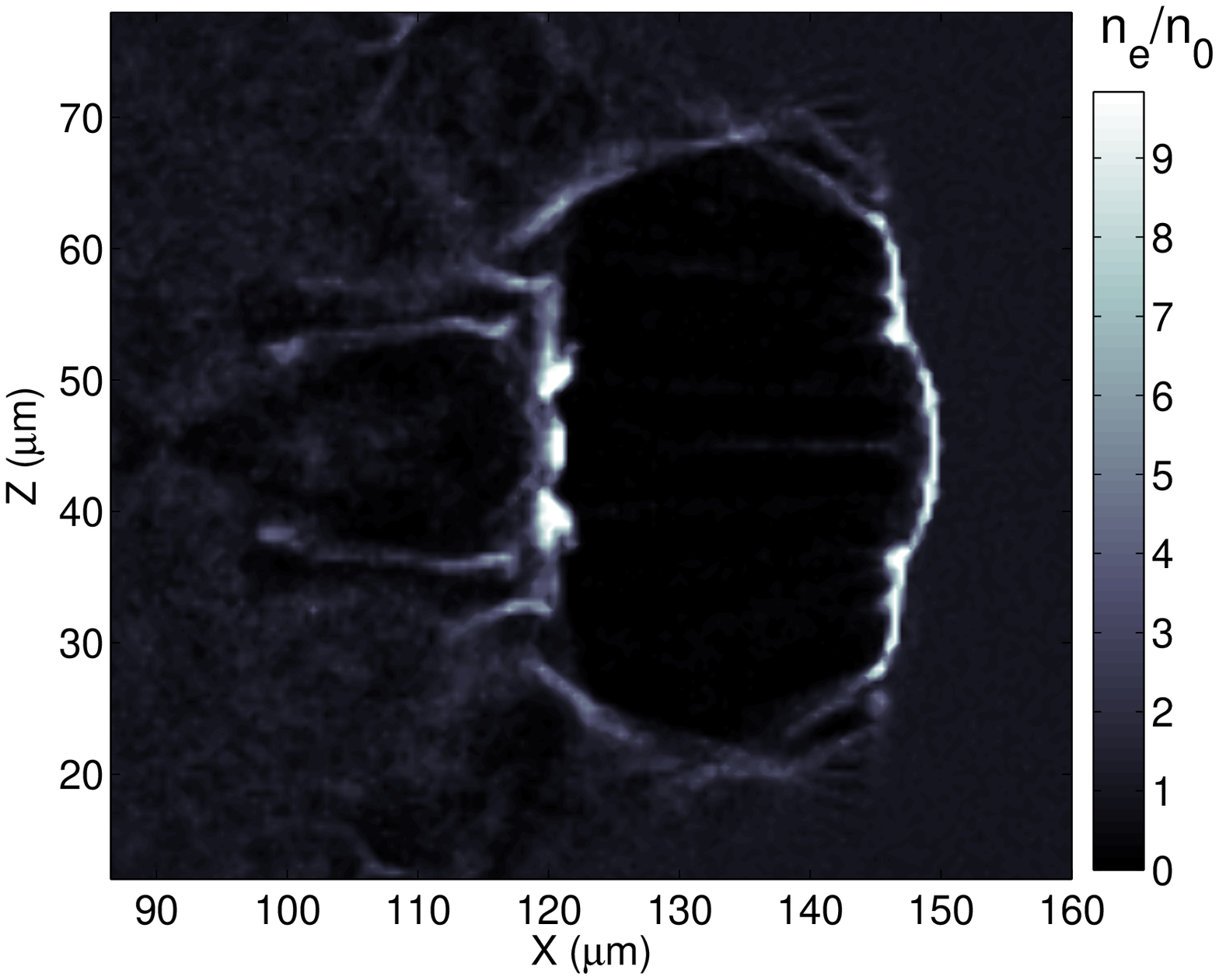}
\includegraphics[width=45mm]{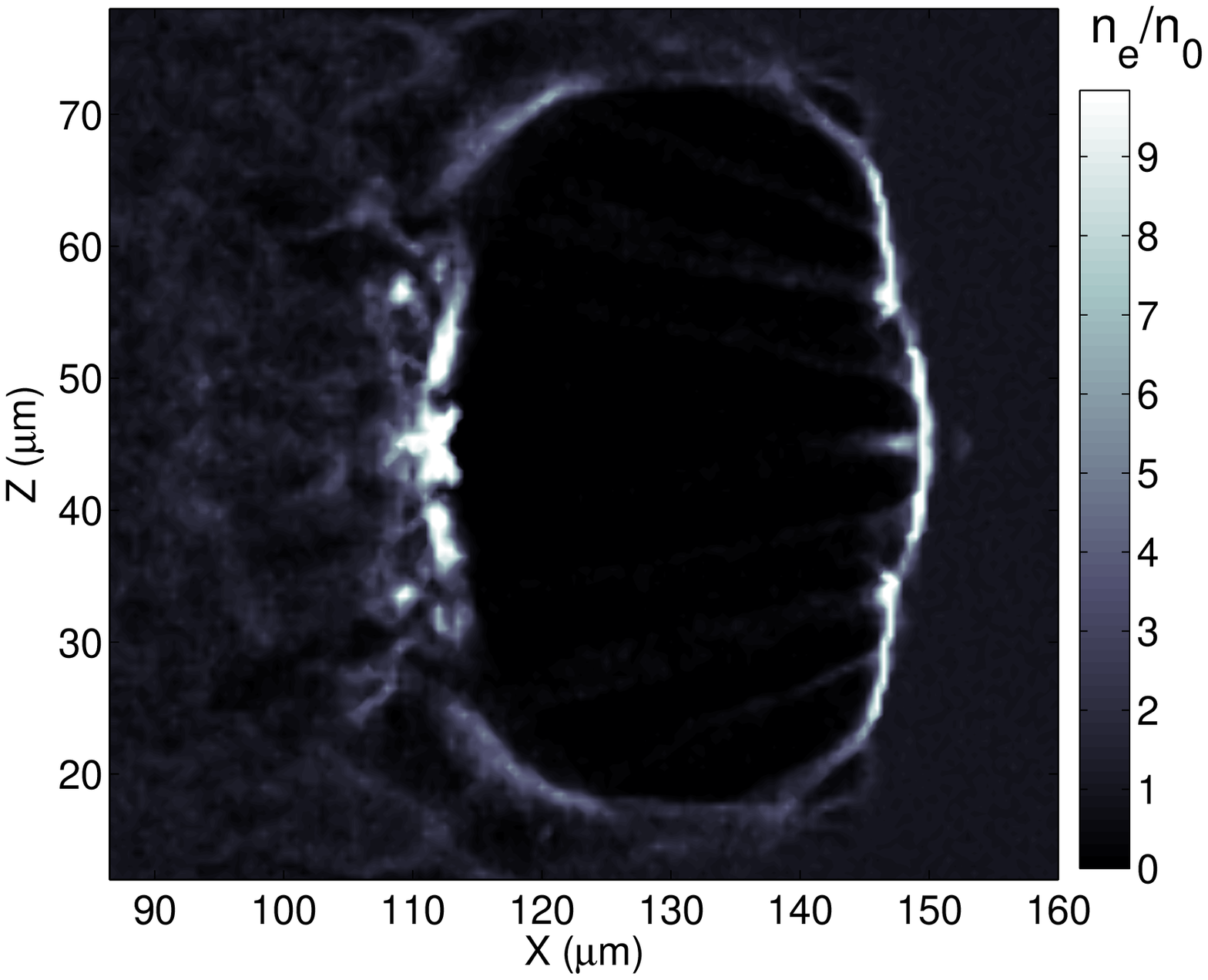}
\caption{ Longitudinal magnetic generated in the upper row and normalized electron density in the lower row observed for the parameter sets shown in Table \ref{simParam}: (a) Sim5, (b) Sim6 and (c) Sim7.  }
\label{BxComp}
\end{figure}

In order to prove the validity of the scalings we show the magnetic field and electron density cross sections from simulation 5, 6, 7 in Fig. \ref{BxComp}, where the laser wavelength is 8 times longer than in Sim1. First Fig. \ref{BxComp}a ($k=0.04$) should be compared with Fig. \ref{Lsp3}a ($k=0.62\times 10^{-3}$). It can be clearly seen that by increasing only the laser wavelength the field amplitude increases by $\approx \sqrt{8}$, and the relative bubble size, $\lambda_p/\lambda_{sp}$, decreases by the same factor: $\approx 2.8$. The bubble can increase again due to the large amount of captured electrons, which decrease the electrostatic potential inside of the bubble, leading to its expansion. The laser intensity and plasma density can be changed such that almost the same magnetic field is obtained (Sim6) and the ratio $\lambda_p/\lambda_{sp}$ becomes larger (Fig. \ref{BxComp}b). Finally, similar bubble shape can be obtained by increasing the laser intensity and decreasing the plasma density (Sim7).

\begin{figure}[h]
\centering

\textbf{(a)}\hspace{40mm}
\textbf{(b)}\hspace{40mm}
\textbf{(c)}

\includegraphics[width=45mm]{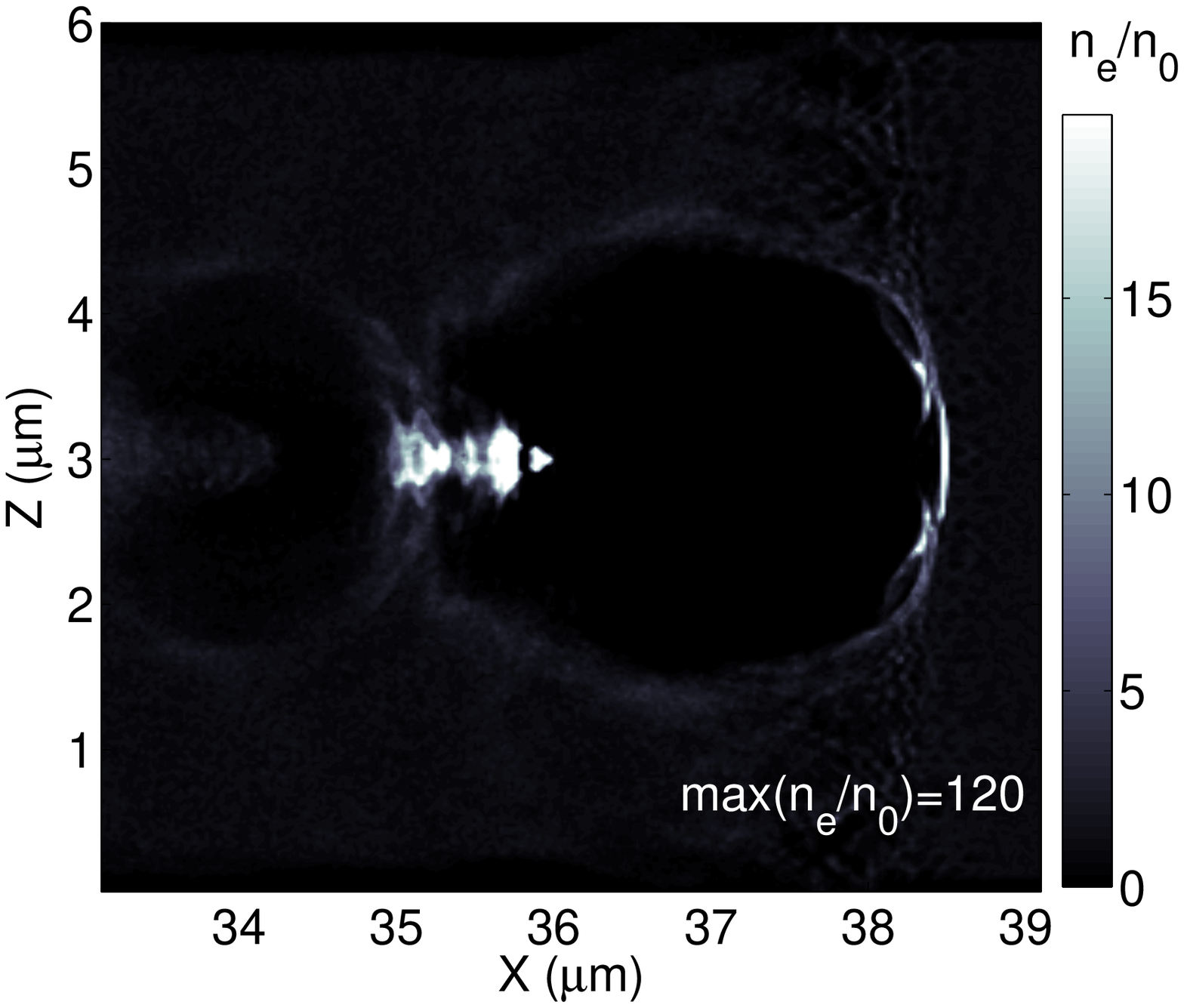}
\includegraphics[width=45mm]{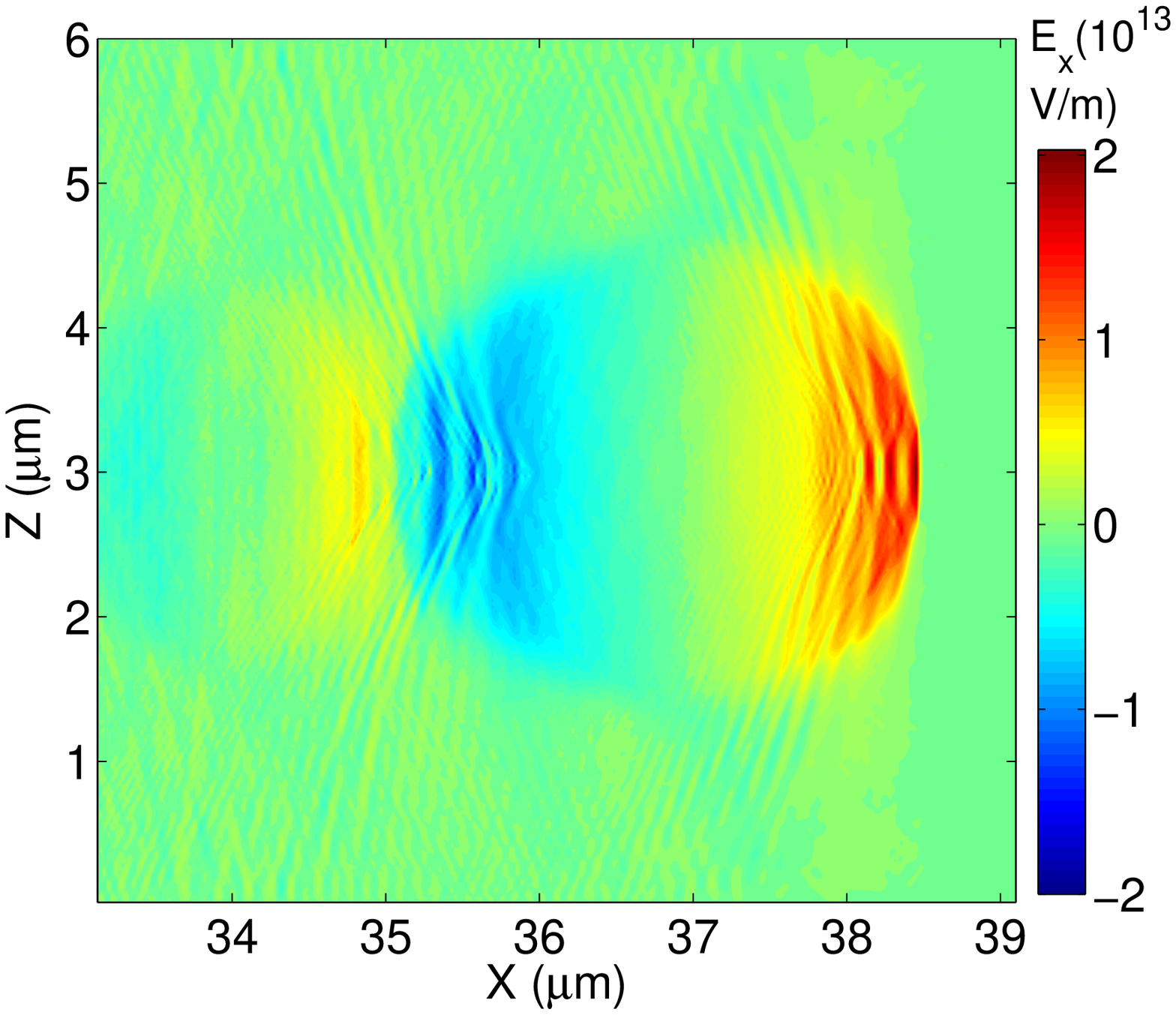}
\includegraphics[width=45mm]{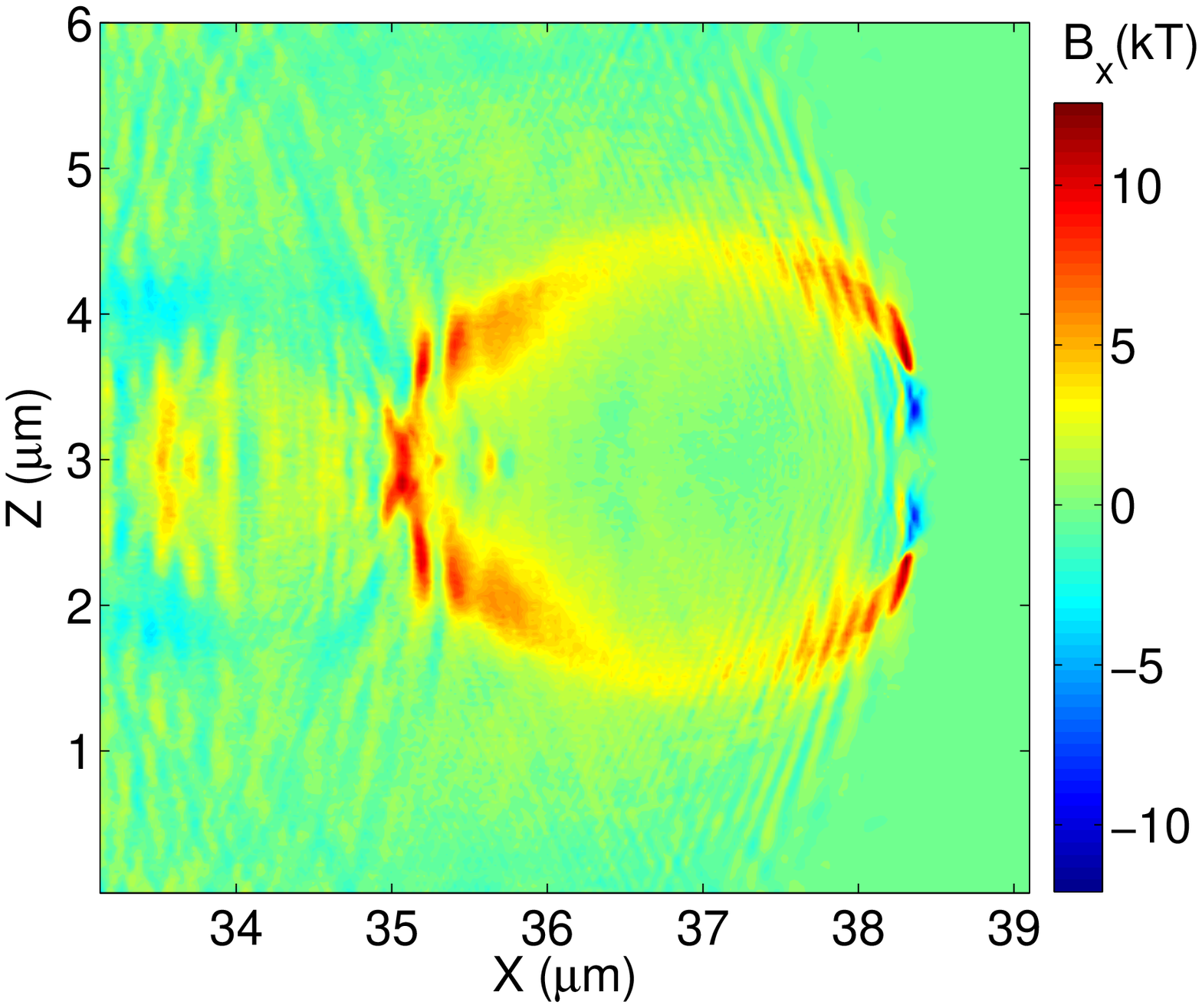}

\textbf{(d)}\hspace{40mm}
\textbf{(e)}\hspace{40mm}
\textbf{(f)}

\includegraphics[width=44mm]{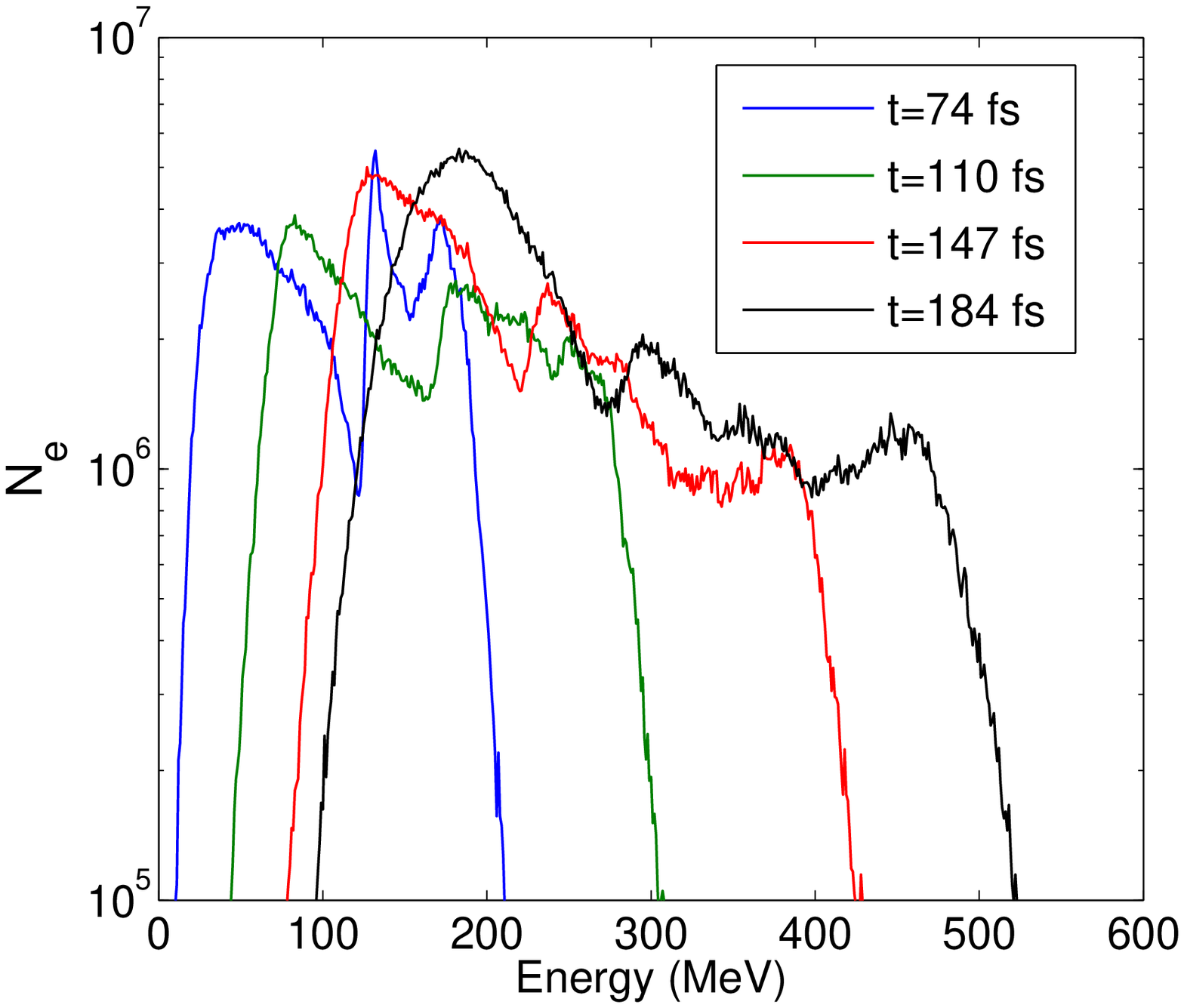}
\includegraphics[width=44mm]{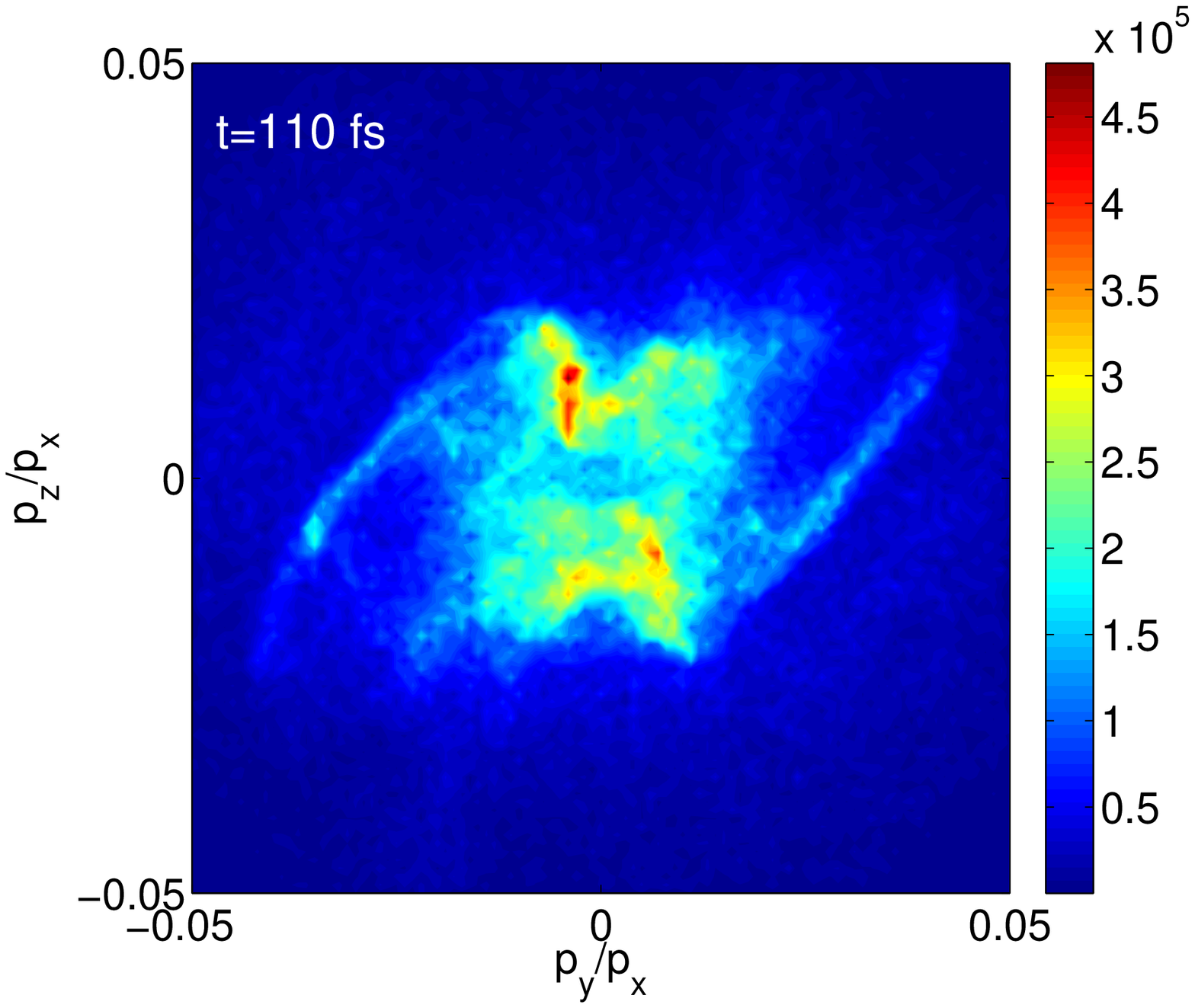}
\includegraphics[width=44mm]{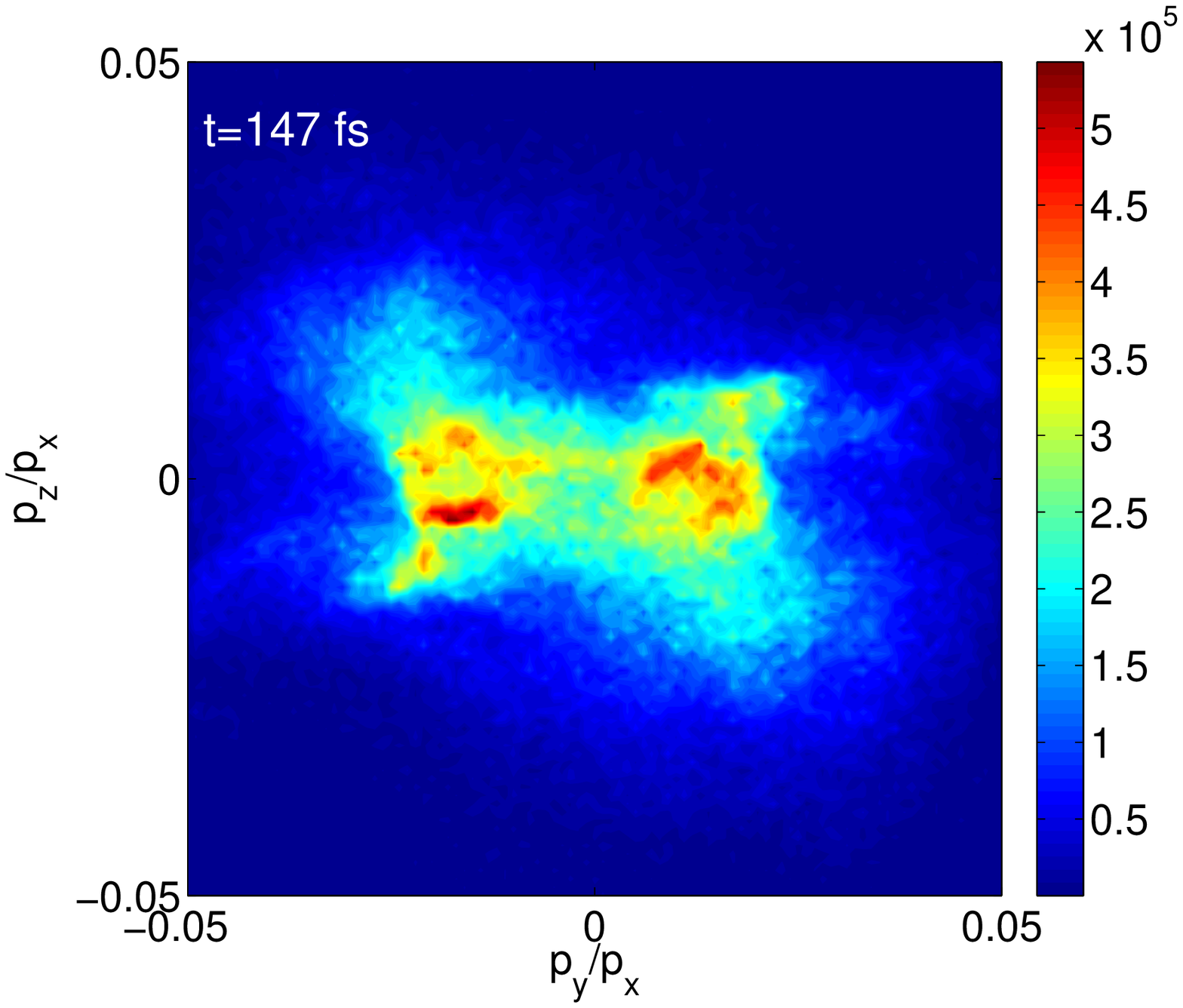}

\caption{ \textbf{(a)} A snapshot of electron density, \textbf{(b)} longitudinal electric field and \textbf{(c)} axial magnetic field  in the case of Sim2 at $t=110$ fs. \textbf{(d)} Electron energy and \textbf{(e,f)} momentum distributions for different time instances. }
\label{elecdistr}
\end{figure}

Plasma waves produced in the under-dense plasmas are famous for their capability of accelerating electrons to GeV energy. This is also true in the case of screw-like-shape of the laser pulse. In this case the longitudinal accelerating field is combined with strong longitudinal magnetic field which is pointing in the same direction giving set of new features to this accelerating scheme. Due to the B-field produced in the tail of the bubble very large number of electrons can be accumulated in the accelerated beam with dramatically improved emittance. In case shown in Fig. \ref{elecdistr} the electron energy has reached the 500 MeV after only 60 $\mu$m of propagation. It is clear that this high energy is due to the high plasma density and high laser intensity used in the simulation. The energy spectrum and angular distribution of electrons is shown in Fig. \ref{elecdistr}d,e,f. Two electron bunches are observable in the density distribution (\ref{elecdistr}a), which appear also in the momentum distribution. Even at this preliminary studies an efficient acceleration and multiple beam bunching is visible. As their energy increases their angular spread decreases and probably will be unified into a single dense and energetic electron bunch if the acceleration is long enough. The energy spread of electrons is relatively large, but the angular spread of the individual bunches is less then 1 mrad. The large energy spread can be attributed to the pulse energy depletion and to the large charge density of the electron beam, which modifies the bubble fields (see Fig. \ref{elecdistr}b,c), making the bunching and acceleration less effective. 

Due to the strong magnetic field these electrons are compressed, and rotate along the propagation axis (see Fig. \ref{elecdistr}e,f). This electron beam dynamics can be very interesting for generating high intensity synchrotron radiation (SR) at $<$ nm wavelength scale predominantly in the forward direction. Since the spatial resolution of our simulation method is not high enough to resolve these wavelengths (and the spiral motion of electrons) this radiation can not be investigated at this stage of the numerical modeling and will be studied in the future. We anticipate that due to electron spiral motion in magnetic field strong energy loss from the transversal motion will take place due to intense SR. The SR losses per unit length are \cite{jackson}:

\begin{equation}\label{eq:enloss}
\frac{dW}{dS}=\frac{2}{3}\frac{r_e \gamma^4 m_e c^2}{R^2},
\end{equation}
where $r_e$ is the classical electron radius and $R$ is the radius of curvature of electron motion defined by the magnetic rigidity, $B\rho[$Tm$]\approx 3.33 p[$GeV/c$]$, and azimuthal angle of electron motion, $\theta$. Representing the spiral motion of relativistic particles with transverse angle $\theta$ in this solenoidal field $B_x$ as $r= \theta \rho \sin (s / \rho)$ where $\rho = B\rho / B_x$, we can find the radius of curvature of this motion as $1/R = d^2 r / d s^2$ which gives us an estimate $R = B\rho /(B_x \theta)$. Let us define the cooling length $L_{cool}$ to be equal to the distance over which the electron will lose all its transverse energy, estimated as $\gamma m_ec^2 \theta^2$, due to SR. Therefore its transverse emittance will decrease by a factor of $e=2.718$ over $L_{cool}$: 

\begin{equation}\label{eq:Lcool}
 L_{cool}=\frac{3}{2}\frac{R^2}{r_e \gamma^3 \theta}=\frac{3}{2} \left(\frac{B\rho}{B_x}\right)^2 \frac{1}{r_e \gamma^3}.
\end{equation}

For a simple estimation one can assume 1 GeV electron energy, which means $\gamma=2000$ and $B\rho=3.3$ Tm, and $B_x=10^5$ T magnetic field resulting in $L_{cool}\approx 60 \mu$m, which is much shorter than the pulse depletion length \cite{depletion}: $L_{pd}\approx 0.45\lambda_p^3/\lambda_L^2 (\lambda_L/1 \mu m) \sqrt{I_0[10^{18} W/cm^2]/1.4} \approx 3.4$ mm in the case of Sim2. This type of effective cooling should improve overall beam emittance as beam will continue accelerate as shown in Fig \ref{elecdistr}(e,f). Thus, the longitudinal magnetic field will acts not only to guide the electron beam (effective beam collimator) but also as a efficient coolant mediating conversion of beam transverse energy into high frequency synchrotron radiation at the same time.

\subsection{The steady solenoid}

So far the regime of large bubble is considered, where laser spot size $W_L<\lambda_p$ and spiral step $\lambda_{sp}<\lambda_p$ were relatively small. In this case the magnetic field moves with the speed of light, constantly appears and disappears in the plasma, which might not be advantageous in some cases and it is difficult to detect with direct methods. If the pulse length approaches or exceeds the bubble length more uniform magnetic field can be observed (see Fig. \ref{Bx_prof2}b or \ref{BxComp}a). In this regime the azimuthal current can be driven resonantly if the plasma wavelength matches the spiral step of the laser pulse. However in this case the bubble cannot be formed and as a result the spiral currents are not limited to a short intervals of bubble shell thus surviving over a longer distance behind the laser pulse. 

\begin{figure}[h]
\centering

\textbf{(a)}
\includegraphics[width=45mm]{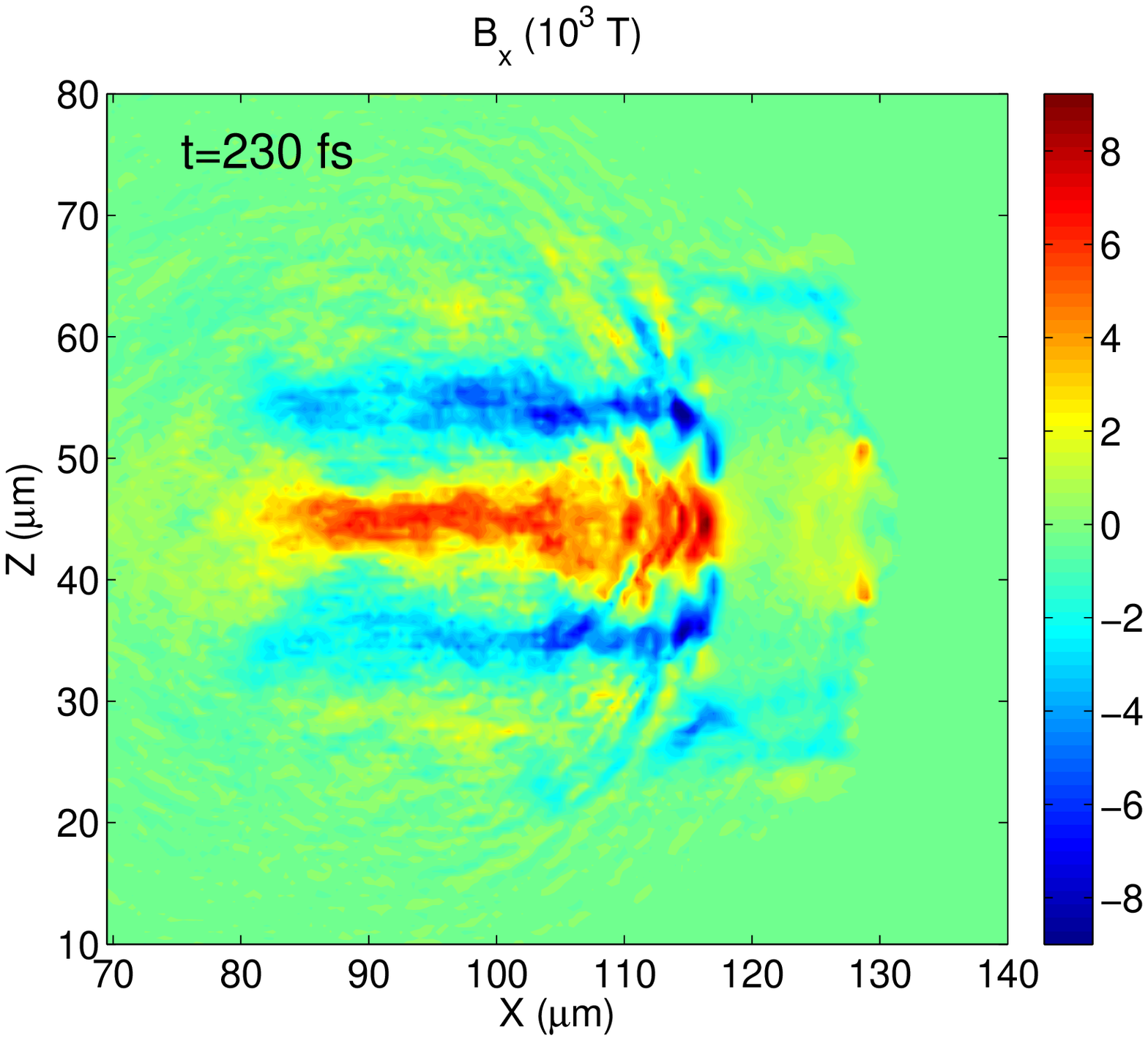}
\includegraphics[width=45mm]{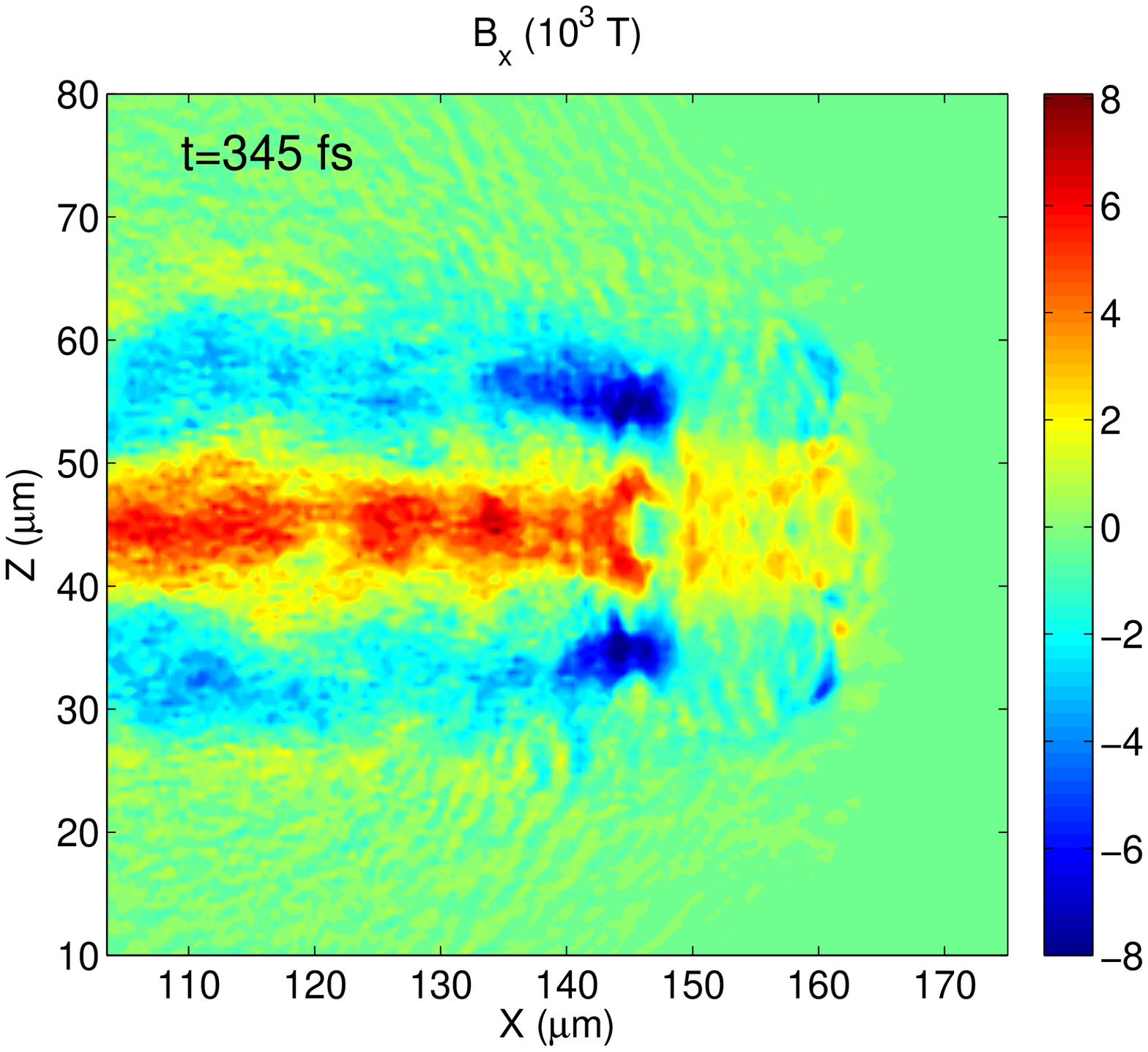}
\includegraphics[width=45mm]{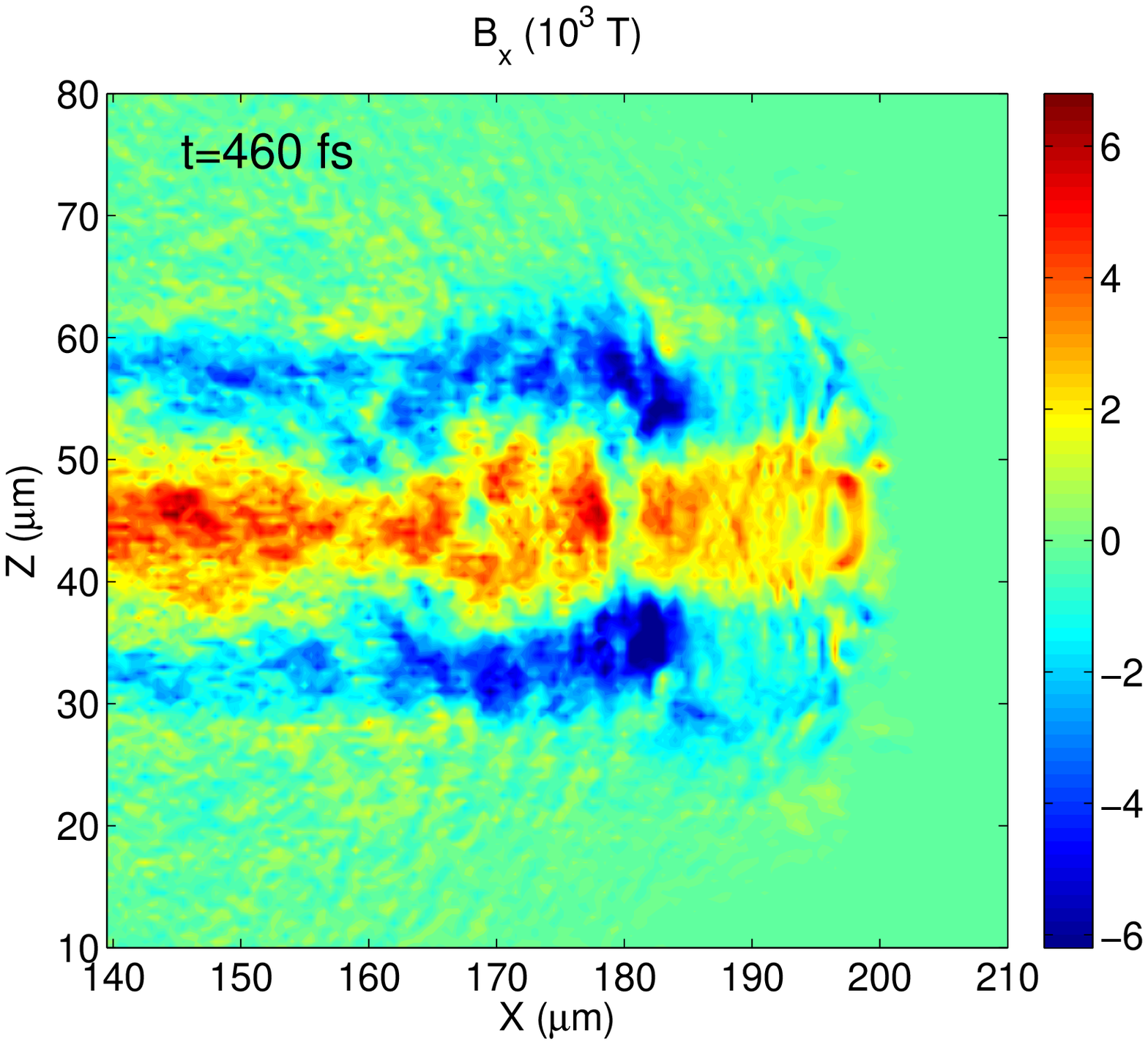}

\textbf{(b)}
\includegraphics[width=45mm]{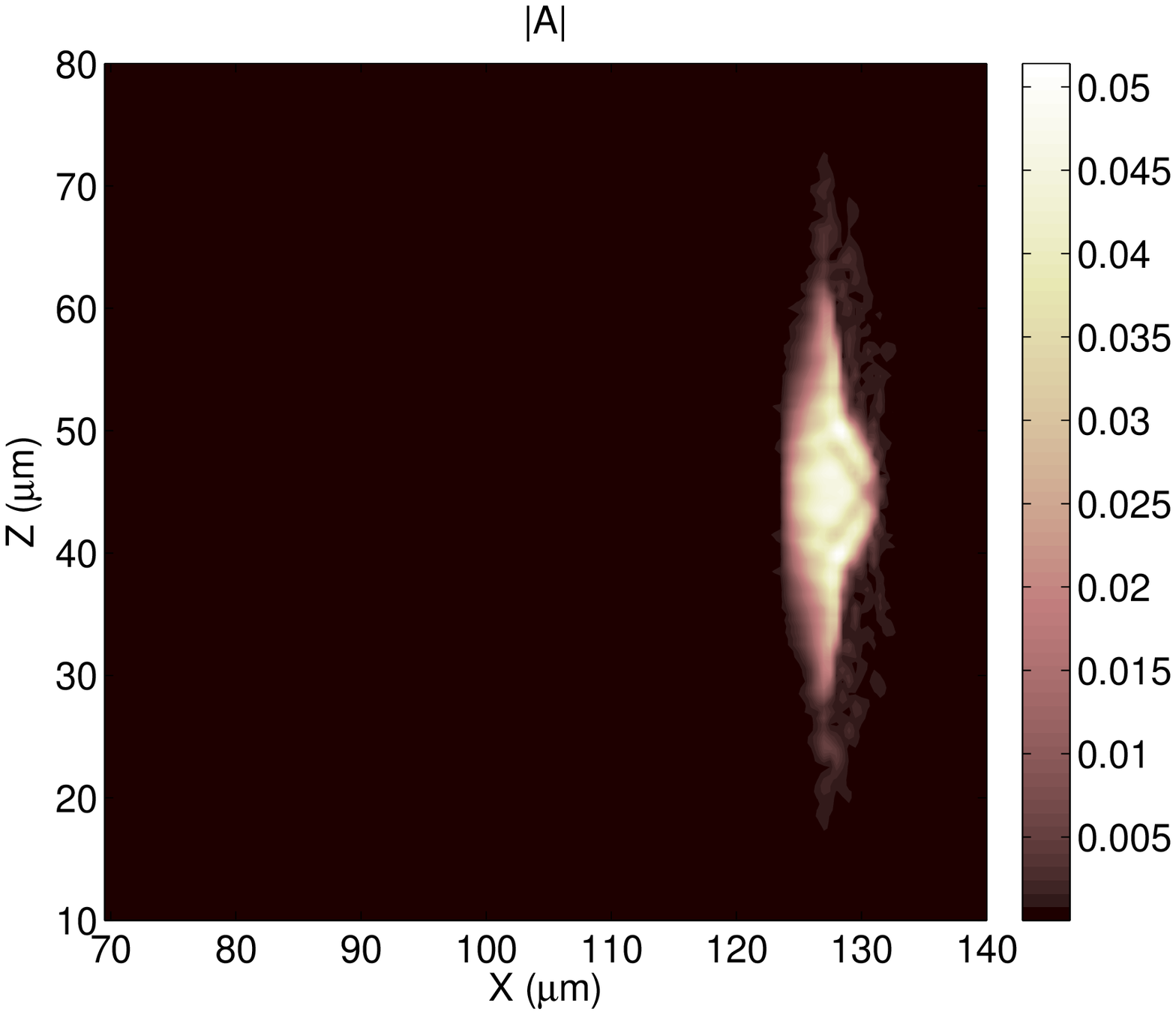}
\includegraphics[width=45mm]{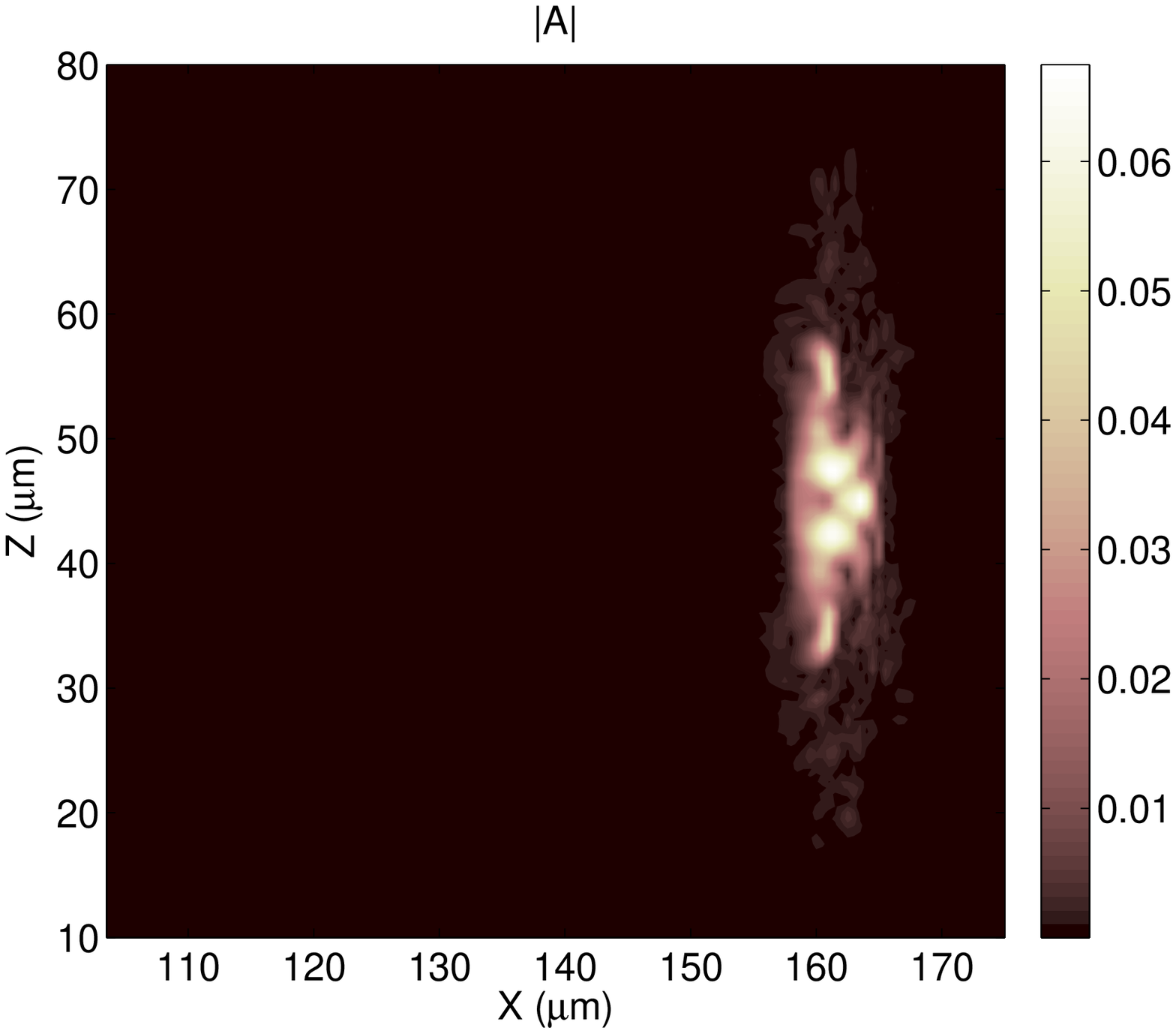}
\includegraphics[width=45mm]{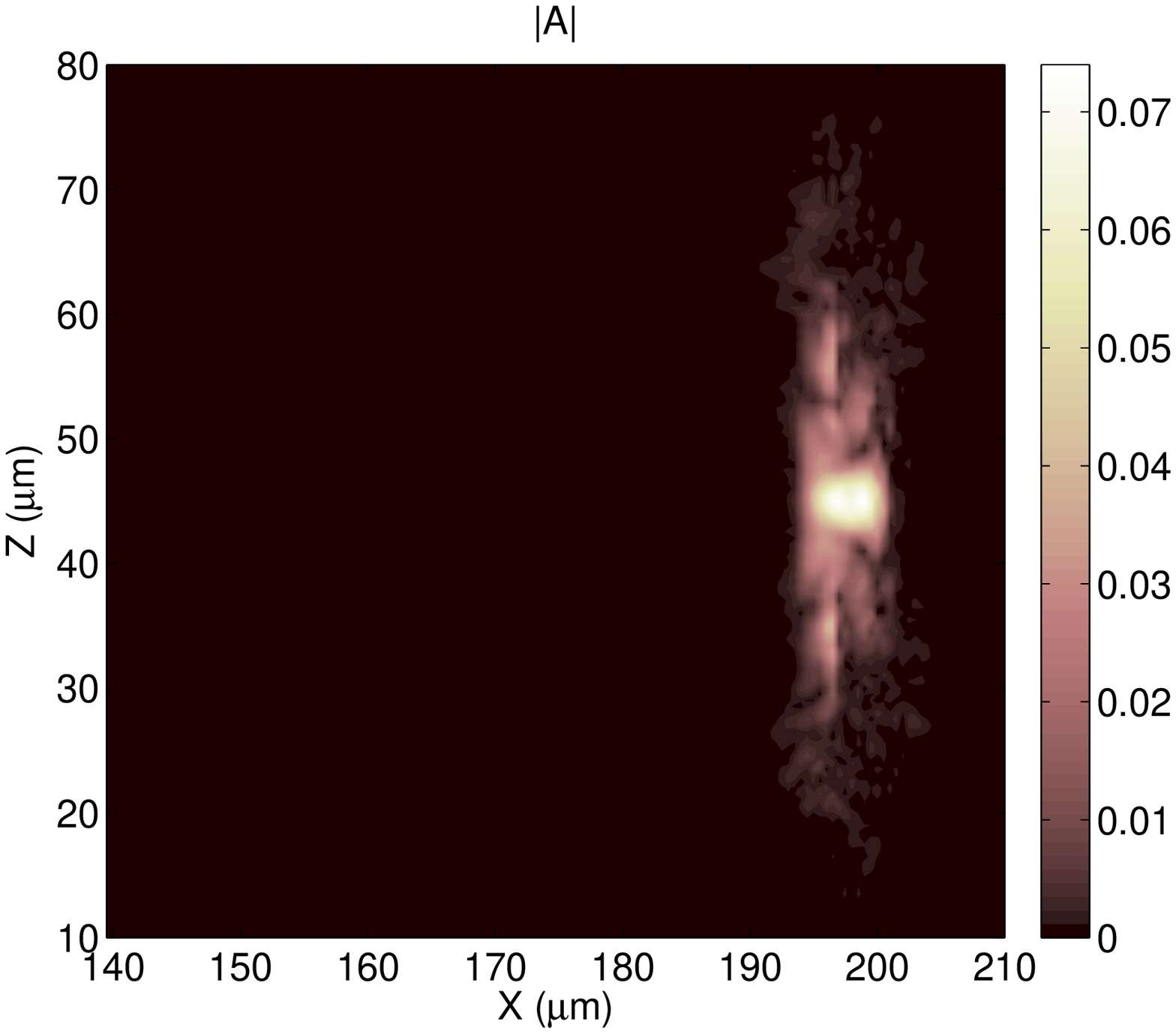}

\caption{ Longitudinal magnetic field {\bf (a)} and amplitude of vector potential in the pulse {\bf (b)} cross sections are shown at three time instances for the parameters of Sim5 with $I_0=0.8\times 10^{21} $W/cm$^2$. }
\label{longB}
\end{figure}

One example is presented in Fig. \ref{longB}, where the parameters are similar to Sim5, but lower intensity is used. Already in Fig. \ref{BxComp}a this regime is observable, because the parameter-set of Sim5 is close to the red area in Fig. \ref{isocurves}, which means fast depletion (decreasing intensity) and transition to the yellow area. The energy distribution of the laser pulse is also shown in Fig. \ref{longB}b, which indicates energy depletion and strong modulations during the pulse evolution. It is clear that all interactions end with transition from bubble to steady solenoid, because of energy depletion. In the channel behind the laser pulse the positive and negative magnetic fields are parallel and separated as expected in any solenoid structures. This structure brakes in the limit of $W_L\gg \lambda_p$, because of magnetic filaments evolving in the plasma if the current flow is wider than the local Debye-length. 

The lifetime of magnetic fields generated in this way depends on the collision frequency between electrons and ions and on the magnetic diffusion time, which is defined as \cite{davies2003} $t_d=\mu_0 R^2/\eta$, where $R$ is the transverse scale length of the electron beam and $\eta$ is the resistivity. The first effect is negligible in gases, but if $>$kT magnetic fields are generated, the plasma density could be high enough to smear out the electron rotation. The second effect is also negligible, because the plasma resistivity is on the order of $10^{-6} \Omega$m, which gives more than 1 ps diffusion time with $R\approx 10 \mu$m. In the simulation shown in Fig. \ref{longB} the depletion time is on the order of 500 fs and the length of the quasi-static axial magnetic field can reach the sub-millimeter level, which can be easily detected in experiments \cite{tatarakis2002}.

\section{Conclusions and outlook}

We suggested and discussed the possibility of generating GigaGauss level axial magnetic field in underdense plasmas in laboratory environment using screw-shaped intense laser pulses. In this paper we consider the pulse length to be equal to the spiral period. In lower density plasmas (gas) the usual $\lambda_L \approx 1 \mu$m high power lasers (available to the researchers nowadays) are suitable for reaching the 10s of kT fields for a short period of time in a spatial volume comparable with the pulse length. It has been demonstrated that by changing the relative laser beam size with respect to the plasma wavelength the shape of the axial magnetic field distribution can be tuned. If the ratio of $\lambda_p/\lambda_{sp}$ is near one, then a static solenoid magnetic field is generated efficiently along a straight line behind the laser pulse limited only by the laser pulse energy depletion. If the plasma wavelength is relatively large a bubble is formed and at its tail a strong, highly peaked and localized magnetic field is generated. This regime becomes highly non-linear (i.e. the introduced scaling does not work) and unstable if the bubble size is too large and the charge of captured electrons influences the longitudinal electric and magnetic fields.

We note that the stable and repetitive production of screw-shape laser pulse still requires large amount of theoretical and experimental work in the field of laser physics. However, we believe that using such a laser pulse can bring a numerous advantages to the fields of laser wake field acceleration and astrophysics. This high magnetic field strength often observed in exotic cosmological objects, which also radiate in a wide range of x-ray and gamma frequency. The observation of radiation in PIC simulations requires very large resolution, which is inadmissible with the current high performance computing units. However, we can anticipate that the spiral motion of well-bunched relativistic electron beam will result in intense synchrotron emission with small angular spread at nm wavelengths and we can make estimates of the corresponding parameters of synchrotron radiation and in particular of the cooling rate of the electron beam emittance. 
  
Immersing electron bunches which are accelerated during laser plasma interaction into co-propagating axial magnetic field enable unique possibility to confine high density beam while improving beam emittance through synchrotron radiation cooling. We have demonstrated that large accelerating potential can be achieved simultaneously with strong longitudinal magnetic field at the point of beam self-injection. Showing that it is possible to use such accelerating structure will motivate future studies in this direction, in particular the investigation of transversal cooling of electrons due to synchrotron radiation, which opens a novel pathway for design of Free Electron Lasers or particle colliders, based on laser plasma acceleration. The non-uniform axial B-field can behave as a magnetic mirror allowing to observe such phenomena as beam reflection and trapping. The possibility of using screw-shaped relativistic electron beams instead of laser pulses is also one of the follow-up research topics triggered by this work.
\newline 

{\bf \large Acknowledgement:}
\newline

We would like to thank the developer team of Tech-X Corporation for the help and support in solving the issues regarding the parallel performance of the simulation code (VSim). The high performance computer cluster was provided by the John Adams Institute (Oxford). The ELI-ALPS project (GOP-1.1.1.-12/B-2012-0001, GINOP-2.3.6-15-2015-00001) is supported by the European Union and co-financed by the European Regional Development Fund.

\newpage

{\bf \large Appendix}
\label{met}
\newline

The cork-screw shape of the laser envelope can be obtained by implementing a rotating elliptic beam profile, with periodicity $\lambda_{sp}$. The transverse beam profile at one point of the pulse is the superposition of two Gaussian shapes: one is rotation symmetric with standard deviation $\sigma_1$ and one is uniform in the radial direction with $\sigma_2 < \sigma_1$ in the orthogonal direction. After some geometric calculations one can derive the function of a line rotating around the axis in the transverse plane, which will give the radial direction of the second Gaussian. Finally, the function describing the intensity distribution has the the following form:

\begin{equation}\label{eq:Int}
I_L=I_0\exp\left(-\frac{D^2}{2\sigma_2^2}\right)\exp\left(-\frac{r^2}{2\sigma_1^2} \right)\cos[(2x/L_{lp}-1)\pi/2],
\end{equation}
where $L_{lp}$ is the total pulse length, $r=\sqrt{y^2+z^2}$ and the distance from the rotating line in the transverse plane:

\begin{equation}\label{eq:distance}
D(x,y,z)=[( y- \cos(\alpha)(z\tan(\alpha)+y))^2 + (z-\sin(\alpha)(z\tan(\alpha) + y))^2]^{1/2},
\end{equation}
where $\alpha=\pi x/\lambda_{sp}$ is the phase of the spiral envelope. In Eq. \ref{eq:Int} a cosine function also appears, which is the longitudinal envelope. The usual Gauss function is not appropriate in this case, because it has to be truncated and the gradient at the leading edge of the pulse would be too large. By using this function we can assure that the intensity value is zero at both ends of the pulse.

The simulation space contains 200 $\times$ 200 $\times$ 200 computational mesh cells with variable grid sizes depending on the bubble size. The propagation time is about 100 fs, which corresponded to about 1000 PIC cycles. The plasma is represented by electrons with under-critical density with 8 macro-particles in each cell. In order to simulate  propagation distance much longer than the domain size we applied the moving window feature of the VSim code. In this case the simulation box is moving in the positive $x$ direction with the speed of light and in each time step electrons are loaded in the new grid cell appearing at the front side and electrons are absorbed at the back side of the simulation domain.


\newpage

\begin{thebibliography}{9}   

\bibitem{zweibel97}
E. G. Zweibel, C. Heiles, Magnetic fields in galaxies and beyond, Nature {\bf 385}, 131 – 136 (1997)

\bibitem{faure04}
J. Faure, Y. Glinec, et al., A laser–plasma accelerator producing monoenergetic electron beams, stimulating coherent radiation, Nature {\bf 431}, p.541 (2004)

\bibitem{remington06}
B. A. Remington, R. P. Drake, D. D. Ryutov, Experimental astrophysics with high power lasers and Z pinches, Rev. Mod. Phys. {\bf 78}, 755 (2006)

\bibitem{elias76}
L. R. Elias, W. M. Fairbank, J. M. J. Madey, H. A. Schwettman, and T. I. Smith, Observation of Stimulated Emission of Radiation by Relativistic Electrons in a Spatially Periodic Transverse Magnetic Field, Phys. Rev. Lett. {\bf 36}, p.717 (1976)

\bibitem{spruit98}
Spruit, H. C. and Phinney, E. S., Birth kicks as the origin of pulsar rotation,  Nature {\bf 393}, 139 (1998)

\bibitem{nature2012}
G. Gregori et al., Generation of scaled protogalactic seed magnetic fields in laser-produced shock waves, Nature {\bf 481}, 480–483 (2012)

\bibitem{balandina12}
Balandina, A.N., et. al., Magnetic Field Generation and Related Topics (MEGAGUSS), 2012 14th International Conference on Megagauss, IEEE 2012, pp.1-7

\bibitem{chernyshev97}
V. K. Chernyshev, V. D. Selemir, and L. N. Plyashkevich, “Super-power explosive magnetic sources for thermonuclear and physical research,” in Megagauss and megaampere pulsed technology and application, Eds. Sarov: RFNC-VNIIEF, pp. 41–58. (1997)

\bibitem{lebedev01}
S. V. Lebedev, et al., X-ray backlighting of wire array Z-pinch implosions using X pinch, Rev. Sci. Instrum. {\bf 72}, 671 (2001) 

\bibitem{turchi80}
P. J. Turchi, A. L. Cooper, R. D. Ford, D. J. Jenkins, and R. L. Burton, “Review of the NRL Liner Implosion Program”, Megagauss Physics and Technology, Ed. Peter Turchi, Plenum Press, New York (1980)

\bibitem{slough12}
J. Slough, A. Pancotti, D. Kirtley, Magnetic Field Generation and Related Topics (MEGAGUSS), 2012 14th International Conference on Megagauss, 2012

\bibitem{sims08}
Sims, J.R. ; Rickel, D.G. ; Swenson, C.A. ; Schillig, J.B. ; Ellis, G.W. ; Ammerman, C.N., Assembly, Commissioning and Operation of the NHMFL 100 Tesla Multi-Pulse Magnet System, IEEE Transactions on Applied Superconductivity  {\bf 18}, p.587-591 (2008)

\bibitem{fujioka2013}
Fujioka et al., Kilotesla Magnetic Field due to a Capacitor-Coil Target Driven by High Power Laser, Sci. Rep. 3 (2013) 1170.

\bibitem{puhkov1996}
Puhkov et al., Relativistic Magnetic Self-Channeling of Light in Near-Critical Plasma: Three-Dimensional Particle-in-Cell Simulation, PRL 76 (1996) 3975

\bibitem{borghesi1998}
Borghesi et al, Megagauss Magnetic Field Generation and Plasma Jet Formation on Solid Targets Irradiated by an Ultraintense Picosecond Laser Pulse, Phys. Rev. Lett. 80 (1998)

\bibitem{tatarakis2002}
Tatarakis et al, Measuring huge magnetic fields, Nature 415 (2002) 280

\bibitem{sarri2012}
Sarri et al, Dynamics of Self-Generated, Large Amplitude Magnetic Fields Following High-Intensity Laser Matter Interaction, Phys. Rev. Lett. 109 (2012)

\bibitem{ali2010}
Ali et al, Inverse Faraday Effect with Linearly Polarized Laser Pulses, Phys. Rev. Lett. 105 (2010) 35001

\bibitem{wang2015}
Wang et al, Hollow screw-like drill in plasma using an intense Laguerre–Gaussian laser, Scient. Rep. 5 (2015) 8274

\bibitem{circularP}
Naseri, Bychenkov, and Rozmus, Axial magnetic field generation by intense circularly polarized laser pulses in underdense plasmas, Phys. Plasmas {\bf 17}, 083109 (2010)

\bibitem{attospiral}
Zs. Lecz and A. Andreev, Attospiral generation upon interaction of circularly polarized intense laser pulses with cone-like targets, Phys Rev E {\bf 93}, 013207 (2016)

\bibitem{helical}
Yin Shi et al., Light Fan Driven by a Relativistic Laser Pulse, Phys Rev Lett {\bf 112}, 235001 (2014)

\bibitem{plasmaRot}
Zs. Lecz, A. Andreev and A. Seryi, Plasma rotation with circularly polarized laser pulse, Laser and Particle Beams {\bf 34}, 31-42 (2015)

\bibitem{amplify}
J. Vieira et al., Amplification and generation of ultra-intense
twisted laser pulses via stimulated Raman scattering, Nature Communications {\bf 7}, 10371 (2016)


\bibitem{envelopeModel}
T. Antonsen and P. Mora, Kinetic modeling of intense, short laser pulses propagating in tenuous plasmas,  Phys. Plasmas {\bf 4}, 217 (1997)

\bibitem{Gordon2000}
D. Gordon, W. Mori, and T. Antonsen, Jr., A Ponderomotive Guiding Center Particle-in-Cell
Code for Efficient Modeling of Laser–Plasma Interactions, IEEE Trans. Plasma Science {\bf 28}, 1135–1143 (2000)

\bibitem{PMessmer}
P. Messmer and D. L. Bruhwiler, Simulating laser pulse propagation and low-frequency wave emission in capillary plasma channel
systems with a ponderomotive guiding center model, Phys. Rev. ST Accel. Beams {\bf 9}, 031302 (2006)

\bibitem{pulseEvol}
C. B. Schroeder, C. Benedetti, E. Esarey, and W. P. Leemans, Phys Rev Lett {\bf 106}, 135002 (2011)

\bibitem{review}
E. Esarey et al., Physics of laser-driven plasma-based electron accelerators,  Review of Modern Physics {\bf 81}, 1229 (2009)

\bibitem{jackson}
J. D. Jackson, Classical Electrodynamics, 3rd ed. (Hamilton, New York, 1999)

\bibitem{davies2003}
J.R. Davies, Electric and magnetic field generation and target heating by laser-generated fast electrons, Physical Review E {\bf 68}, 056404 (2003)

\bibitem{depletion}
Shadwick, B. A., C. B. Schroeder, and E. Esarey, Physics of Plasmas {\bf 16}, 056704 (2009)


\end{thebibliography}
\end{document}